\def\'#1{\ifx#1i{\accent"13 \i}\else{\accent"13 #1}\fi}
\newcommand{\arcsec}{\hbox{$^{\prime\prime}$}}

\newcommand{\degr}{\hbox{$^\circ$}}

\newcommand{\Hbeta}{{H$\beta$}}
\newcommand{\kms}{{km\,s$^{-1}$}}

\newcommand{\Vdiez}{$V_{10}$}
\newcommand{\Vexp}{$V_{\mathrm{exp}}$}

\documentclass{rmaa}

\title{Galactic Planetary Nebulae with Wolf-Rayet Nuclei III. \linebreak
 Kinematical Analysis of a Large Sample of
Nebulae\altaffilmark{1} \medskip }

\author{
S. Medina,\altaffilmark{2}
M. Pe\~na,\altaffilmark{2}
C.~Morisset,\altaffilmark{2} and
G.~Stasi\'nska\altaffilmark{3}}

\altaffiltext{1}{Based upon data collected at the Observatorio
Astron\'omico Nacional, San Pedro M\'artir, B.~C., M\'exico.}
\altaffiltext{2}{Instituto de Astronom\'ia, Uni\-ver\-sidad Nacio\-nal Aut\'o\-noma de M\'exico,
M\'exico.}
\altaffiltext{3}{LUTH, Observatoire de Meudon, France.}

\shortauthor{Medina et~al.}
\shorttitle{Galactic PN with Wolf-Rayet Nuclei}

\fulladdresses{
\item Selene Medina,  Christophe Morisset, and
Mirian Pe\~na: Instituto de Astronom\'{\i}a, UNAM,
       Apdo. Postal 70-264, 04510 M\'exico, D.~F., M\'exico (miriam,\,morisset,\,selene@astroscu.unam.mx).
\item Gra\.zyna Stasi\'nska: LUTH, Observatoire de Meudon, 5, place Jules Janssen, 92195 Meudon Cedex, France \linebreak (\email{grazyna.stasinska@obspm.fr}).
}
\listofauthors{S.~Medina, M.~Pe\~na, C.~Morisset, \& G.~Stasi\'nska}
\indexauthor{Medina,~S.}
\indexauthor{Pe\~na,~M.}
\indexauthor{Morisset,~C.}
\indexauthor{Stasi\'nska,~G.}

\abstract{
Expansion velocities (\Vexp) of different ions and line widths at the base of the lines are 
measured and analyzed for 24 PNe with [WC]-type nuclei (WRPNe), 9 PNe ionized by WELS 
(WLPNe) and 14 ordinary PNe. A comparative study of the kinematical behavior of the sample clearly 
demonstrates that WRPNe have in average 40-45\% larger \Vexp, and possibly more turbulence than 
WLPNe and ordinary PNe. WLPNe have velocity fields very much alike the ones of ordinary PNe, rather 
than the ones of WRPNe. All the samples (WRPNe, WLPNe and ordinary PNe) show expansion velocities 
increasing with age indicators, for example $<$\Vexp$>$  is larger
for low-density nebulae and also it is larger for nebulae around high-temperature stars. This age effect is much stronger for evolved WRPNe, suggesting that the [WC] winds 
have been accelerating the nebulae for a long time, while for non-WRPNe the acceleration seems to
stop at some point when the star reaches a temperature of about 90\,000-100\,000 K. Non-WR nebulae reach a maximum \Vexp~$\leq $~30 \kms \ while evolved WRPNe reach maximum \Vexp \ of about 40 \kms. For all kind of objects  (WRPNe and non-WRPNe) it is found that in average \Vexp(N$^+$) is 
slightly larger than \Vexp(O$^{++}$), indicating that the nebulae present  acceleration of the 
external shells.}

\resumen{}

\addkeyword{ISM: Kinematics and dynamics}
\addkeyword{Planetary nebulae: general}
\addkeyword{Planetary nebulae: individual: NGC\,6369, M\,1-32, BD+30\degr3639, K\,2-16}
\addkeyword{Stars: Wolf-Rayet}

\begin{document}
\maketitle

\section{Introduction}
Among galactic planetary nebulae with known central stars, less than 15\%
 have nuclei presenting Wolf-Rayet features.  All these stars 
have been classified as belonging to the [WC] spectral type, showing almost pure
helium and carbon in their atmospheres (e.g.,  Tylenda et al. 1993; Hamann 1997).
The only possible [WN]  central star in the Galaxy was reported by Morgan et al. (2003). 

Many recent studies have been devoted to these planetary nebulae (hereinafter
WRPNe) and their central stars with interesting new results.  G\'orny \&
Stasi\'nska (1995) found that the physical properties and chemical abundances of
WRPNe were not different from those of non-WRPNe, except for higher expansion
velocities.  Since then, Pe\~na et~al.\@(1998, Paper~I) found that central stars 
of different initial masses can pass through the same [WC] stage and Pe\~na, Stasi\'nska, \& Medina 
 (2001, Paper~II) reported an unexpected behaviour
of the [\ion{O}{III}]/[\ion{N}{II}] electron temperature ratios for WRPNe which
does not seem to be present in non-WRPNe.  In addition, G\'orny et~al.\@  (2001)
found that in an infrared diagram (\textit{H--K}) versus (\textit{J--H}),
WRPNe spread over a larger zone than non-WRPNe which is probably more
reflecting differences in dust and stellar properties than differences in
nebular properties.  Acker et~al.\@ (2002) have concluded that WRPNe show similar 
expansion velocities but larger turbulent velocities that normal PNe,
and recently Gesicki et al. (2003) concluded that the WRPNe tend to 
show strong acceleration at the ionization front and strong turbulent motions.

For the central stars, several scenarios to produce such H-deficient low-mass
stars have been proposed (e.g.,  Bl\"ocker 2001; Herwig 2001).  In addition,
G\'orny \& Tylenda (2000) and De Marco (2002) have provided further arguments 
for the existence of the evolutionary sequence: [WC]-late $\rightarrow$
[WC]-early  stars, ending with the PG~1159 type stars, proposed by
Acker et al.  (1996) and Hamann (1997).  However, this proposition 
is still debated (Hamann et al. 2005).
Parthasarathy et al. (1998) have claimed that ``weak emission line stars''
(WELS), as defined by Tylenda et~al.\@ (1993), are an intermediate
stage between [WC] and PG~1159 stars.  However, Pe\~na et~al.\@ (2001) have argued 
that PNe around WELS are more similar to ordinary PNe than to WRPNe and consider
doubtful the proposition of Parthasarathy et~al.

Since 1995 we have performed systematic observations of WRPNe, to
gather a homogeneous high-resolution spectroscopic data set for studying the
nature and evolutionary status of WRPNe.  From these data, in Paper~I we presented 
detailed photoionization models for a sample of very high 
excitation nebulae ionized by [WC 2-3] stars, and  in Paper~II we reported 
line intensities, physical conditions and chemical composition 
for 34 PNe with nuclei of all [WC] spectral types, including a few WELS.

In this work, our data are employed to analyze the kinematical
behaviour of 24 WRPNe in order to  study the effects of
the [WC] winds on the nebular shells.  We analyze some 
phenomena affecting the nebular profiles such as expansion velocities,
turbulence and high velocity material.  In particular we address the problem of whether 
WRPNe present higher expansion velocities and/or turbulence than non-WRPNe.  For our analysis we have included a sample of 14 ordinary planetary nebulae
(i.e., whose nuclei are neither WR nor WELS) which are useful as a control sample to confirm
that any special trend found for WRPNe is not an instrumental effect.  Nine
PNe ionized by a  WELS (hereinafter WLPNe) have been included also in the analysis.

In \S~2, the observations and data reduction are described.  The line
profiles are discussed in \S~3.  The expansion velocities derived for the
sample are presented and analyzed in \S~4, where a brief discussion
regarding turbulence can be found.  In \S~5, we present the line widths at the 
base of lines, searching for the highest velocity material. The kinematical analysis 
of selected nebular lines is presented in \S~6.  Our main conclusions are summarized 
in \S~7.  Finally,  two Appendices were included where  we present some 
examples of our data and discuss some objects with particularly interesting 
line profiles.

\begin{table}[!h]\centering
  \setlength{\tabnotewidth}{\columnwidth}
 \setlength{\tabcolsep}{1.8\tabcolsep}
  \tablecols{8}
\caption[]{Observing log.\tabnotemark{a}}
\label{Tabla 1}
\begin{tabular}{clccclcc}
\toprule
PN G & Usual name & Obs. dates \\
&  & (dd/mm/yy)  \\
\midrule
$013.7-10.6$ & YC 2-32      & 02/11/00 \\
$037.7-34.5$ & NGC 7009     & 01/11/00 \\
$084.9-03.4$ & NGC 7027     & 02/11/00 \\
$103.7+00.4$ & M 2-52       & 02/11/00 \\
$104.4-01.6$ & M 2-53       & 01-02/11/00 \\
$118.0-08.6$ & Vy 1-1       & 02/11/00 \\
$130.3-11.7$ & M 1-1        & 01-02/11/00 \\
$133.1-08.6$ & M 1-2        & 01-02/11/00 \\
$147.4-02.3$ & M 1-4        & 03/11/00 \\
$159.0-15.1$ & IC 351       & 11/12/98 \\
$166.1+10.4$ & IC 2149      & 13/12/98 \\
$194.2+02.5$ & J 900        & 02/11/00 \\
$196.6-10.9$ & NGC 2022     & 02/11/00 \\
$215.2-24.2$ & IC 418       & 12-13/12/98 \\
$221.3-12.3$ & IC 2165      & 03/11/00 \\
$243.8-37.1$ & PRTM 1       & 12-13-14/12/98 \\
$294.1+43.6$ & NGC 4361     & 04/10/99 \\
\bottomrule
\tabnotetext{a}{Observing set-up: $2048 \times 2048$ ($14 \times 14$\,$\mu{\mathrm{m}}^2$) pixel CCD detector;
slit dimensions: ${4''} \times {13''}$, oriented E--W; wavelength range 3360--7360\,\AA,  spectral resolution $\sim 0.2$\,\AA. Extraction window: ${4''}\times{3''}$, for all the objects.}
\end{tabular}
\end{table}

\section{Observations and data reduction}
 High-resolution spectroscopic data were systematically gathered with the 2.1\,m 
telescope at the Observatorio Astron\'omico Nacional, San Pedro 
M\'artir, B.~C., M\'exico on 1995 July 29--31, 1996 June 14--17, 1997 August 3--4, 1998, 
December 11--14, 1999 October 4--7, and 2000 November 1--3.  The REOSC Echelle 
Spectrograph plus the University College London camera (described by Levine \& 
Chakrabarty 1993) were always used.  Two different CCD detectors were employed:  
for observations on 1995 and 1996 we used a CCD Tektronix of $1024 \times 1024$ 
(23$\mu$m) pixels, providing an average spectral resolution of 0.3\,\AA \ (about 
18\,\kms \ in velocity) with a 150$\mu$m slit width, and since 1997 we used a 
CCD Thomson of $2048 \times 2048$ (14$\mu$m) pixels, giving a 
resolution of 0.2\,\AA \ (about 11\,\kms) with the same slit width. 

The observations for most of the objects presented here, were described in Paper 
II, where a detailed list of objects, observing dates, exposure times and instrumental 
set-up were given.  In Table~1 we present the log of observations for objects not 
reported in Paper~II. As usual we have taken at 
least two consecutive observations for each object. Different exposure 
times (from 2 to 15 minutes) were used to obtain a S/N  better than 3 for the 
weak lines without saturating the most intense ones.  The different exposures 
for the same object were averaged to increase the S/N.  The slit was always E--W oriented.  
In this work, the reported data correspond always to those obtained for the central zone.  For all the objects we covered a wide spectral range (at least from 3700 to 6800 \AA),
therefore lines for high and low ionized species were obtained.  This is important
for studying the kinematics on different zones of the nebula where such species reside.

A Th--Ar lamp was used for wavelength calibration in all the spectral ranges and standard stars from the list by Hamuy
et~al.\@ (1992) were observed each night for flux calibration.  Data reduction was
performed using the package IRAF\footnote{IRAF is distributed by NOAO, which is
operated by AURA, Inc.  under contract with the NSF.}  and includes standard
bias-subtraction and flat-field correction for all spectra.

In Table~2, we present some nebular characteristics (diameter, electron temperatures 
from the [\ion{O}{III}] lines, densities from the [\ion{S}{II}] lines and ionic abundance 
ratios) for all the objects for which the kinematical analysis is performed.    
Most of the data  was taken from Papers~I and II.  For the new objects (Table~1) the nebular
characteristics were derived by using the same procedures as reported in those
papers.  In addition, in columns 8 to 11 we present the stellar classification,
stellar temperature and wind parameters, taken from the literature.  All these data
will be used in the interpretation of the kinematical data.



\section{Line profiles} 

 [WC] stars present intense and wide emission lines of He, C, and other 
elements, whose wavelengths occasionally coincide  with the wavelengths of some nebular 
lines. For instance, this is the case of the nebular \ion{He}{II}, \ion{He}{I}, and 
H Balmer lines which coincide with stellar  \ion{He}{II} and \ion{He}{I} lines, and 
especially it is the case of  [\ion{N}{II}]~$\lambda$6583  and [\ion{S}{II}]~$\lambda$6731 
which appear on the top of some \ion{C}{II} multiplet 2 lines. High spectral 
resolution is necessary to deblend the nebular and stellar components and to 
study the line profiles.  This is  
essential for compact nebulae around [WC]-late stars where the stellar 
winds  show terminal velocities of a few 100 \kms, and the stellar lines 
present a FWHM of a few \AA. Therefore  a spectral resolution  better than 
5000 is necessary. This is one of the advantages of our spectra (resolution of about 20,000).  
In Appendix A, we present stellar and nebular lines for some of our objects to show that our spectral resolution allows 
us to safely deblend both components for most of the objects. Thus the nebular lines 
have been measured without stellar contamination.

Also our high spectral resolution, allows us to obtain well resolved nebular 
profiles for all the sample, except for a couple of objects showing FWHM of lines
smaller than our instrumental resolution.  Thus, we have an extensive data set
including different kind of objects, obtained in a consistent way which allows us to
perform a reliable comparative analysis of the kinematics. First we will analyze the 
adequacy of deriving expansion velocities from the line profiles.

\subsection{Line Profiles and Velocity Field}

A nebular line profile depends on several internal properties such as the nebular
morphology, the density, thermal and ionization structures, the expansion and turbulence  velocities, etc.  In addition, the observed profile also
depends on the spectral resolution and the position and dimensions of the slit: 
for instance, when the nebula is extended and the slit just covers the central zone, 
the classical double-peak profile is obtained.  Many authors have derived the 
nebular expansion velocity, \Vexp, as half the separation of the peaks.  On the other 
hand, when the slit includes the entire nebula, or we are dealing with a filled nebula 
(with a very small inner hole) a single 
profile is usually observed.  In such a case, the half width at half maximum (HWHM) has 
been used to measure \Vexp.  In our sample, many nebulae do not present a well defined
shell, or they have knots at different velocities, showing then complex profiles, for 
which is difficult to derive \Vexp\ in a straightforward manner.

\begin{table*}[]\centering
\setlength{\tabnotewidth}{0.87\textwidth}
 \setlength{\tabcolsep}{0.5\tabcolsep}
\tablecols{12}
\caption{Nebular and stellar physical parameters\tabnotemark{a}}
\footnotesize{
 \begin{tabular}{lcccccccrccl}
\toprule
PN G & Name & $T_e$  & $N_e$ & He$^{++}$/He$^+$ & O$^{++}$/O$^+$ & $\phi$ & [WC]\tabnotemark{b} & $T_\ast$ & v$_\infty$ & log $\dot M $ & Ref. \\
     &  & (10$^4$\,K)   & (cm$^{-3}$) &   & & (arcsec) & & (10$^3$\,K)  & (km\,s$^{-1}$) &(M$_\odot$\,yr$^{-1}$)  &  \\
\midrule
$001.5-06.7$ & SwSt 1         & 1.06  & 30000  & 0.0    & 0.3   & 5 & 9   & 40    & 900  &  $-6.72$    & dM01  \\
$002.4+05.8$ & NGC 6369       & 1.01  & 3220   & 0.1    & 15.7  & 38 & 4   & 150   & 1200 &  $-6.15$    & KH97a \\
$002.2-09.4$ & Cn 1-5         & 1.09  & 4700   & 0.0    & 82.8  & 7 & 4   & $<57$ & \nodata &  \nodata & TS94  \\
$003.1+02.9$ & Hb 4           & 0.96  & 4600   & 0.1    & 33.9  & 7.2 & 3-4 & 86    & \nodata &  \nodata & TS94  \\
$004.9+04.9$ & M 1-25         & 0.80  & 10400  & 0.0    & 43.6  & 3.2 & 6   & 60    & \nodata &  \nodata & L96   \\
$006.8+04.1$ & M 3-15         & 0.87  & 7443   & 0.0    & 38.8  & 4.5 & 5   & 55    & \nodata &  \nodata & ZK93  \\
$011.9+04.2$ & M 1-32         & 0.99  & 6720   & 0.0    & 12.1  & 9 & 4-5 & 66    & \nodata &  \nodata &  TS94 \\
$012.2+04.9$ & PM 1-188      & 0.94   & 200    & 0.0    & 0.03  & 4 & 10  & 35    & 360  &  $-5.70$ & LH98  \\
$017.9-04.8$ & M 3-30         & 1.03  & 525    & 0.9    & 27.2  & 20 & 2 & 97    & \nodata & \nodata & TS94  \\
$027.6+04.2$ & M 2-43         & 1.05  & 10100  & 0.0    & 14.3 & 1.5 & 8   & 65    & 850  &  $-6.08$ & LH98  \\
$029.2-05.9$ & NGC 6751       & 1.12  & 2730   & 0.0    & 5.2  & 20.5 & 4   & 135   & 1600 &  $-6.12$ & KH97b \\
$048.7+01.9$ & He~2-429       & 0.77  & 6810   & 0.0    & 3.6 & 4.2  & 4-5 & \nodata & \nodata &  \nodata &       \\
$061.4-09.5$ & NGC 6905       & 1.21  & 1530   & 2.9    & 58.6 & 40 & 2-3 & 141   & 1800 &  $-6.32$ & KH97a \\
$064.7+05.0$ & BD+30\degr3639 & 1.01  & 21200  & 0.0    & 0.0  & 7.7 & 9   & 47    & 700  &  $-4.87$ & L96   \\
$068.3-02.7$ & He 2-459       & 2.00: & 17400  & 0.0    & \nodata & 1.3 & 8   & 77    & 1000 &  $-5.01$ & L97   \\
$089.0+00.3$ & NGC 7026       & 0.92  & 3250   & 0.2    & 14.7 & 20 & 3   & 130   & 3500 &  $-6.34$ & KH97a \\
$096.3+02.3$ & K 3-61         & 0.89  & 1690   & 0.0    & \nodata & 6.1 & 4-5 & \nodata & \nodata & \nodata &      \\
$120.0+09.8$ & NGC 40         & 1.00  & 1980   & 0.0    & 0.1  & 48 & 8   & 78    & 1000 &  $-5.62$ & L96   \\
$130.2+01.3$ & IC 1747        & 1.12  & 2380   & 0.2    & 78.6  & 13 & 4   & 126   & 1800 &  $-6.58$ & KH97a \\
$144.5+06.5$ & NGC 1501       & 1.05  & 1020   & 0.7    & \nodata & 52 & 4   & 135   & 1800 &  $-6.28$ & KH97a \\
$146.7+07.6$ & M 4-18         & 0.86  & 6350   & \nodata & 0.0 & 3.7 & 10  & 31    & 160  &  $-6.05$ & dMC99 \\
$161.2-14.8$ & IC 2003        & 1.28  & 3840   & 1.38   & 21.4 & 8.6 & 3   & 88    & \nodata & \nodata & TS94  \\
$243.3-01.0$ & NGC 2452       & 1.30  & 1590   & 0.7    & 6.0  & 19 & 2   & 141   & 3000 &  $-6.20$ & KH97a \\
$352.9+11.4$ & K 2-16         & 1.17  & 504    & \nodata & 0.4  & 13.5 & 11  & 30    & 300  &  $-6.36$ & L97b  \\
$009.4-05.0$ & NGC 6629       & 0.84  & 3470   & 0.0    & 13.5 & 15.5 & wl  & $<52$ & \nodata & \nodata & TS94  \\
$010.8-01.8$ & NGC 6578       & 0.79  & 2270   & 0.0    & 240  & 8.5 & wl  & 65    & \nodata & \nodata & TS94  \\
$011.7-00.6$ & NGC 6567       & 0.95  & 4360   & 0.01   & 28.3 & 7.6 & wl  & 61    & 1950 & \nodata & TS94  \\
$096.4+29.9$ & NGC 6543       & 0.79  & 5640   & 0.0    & 75.4 & 19.5 & wl  & $<66$ & 1900 & $-7.4$ & TS94  \\
$100.6-05.4$ & IC 5217        & 1.12  & 8850   & 0.1    & 93.9 & 6.6 & wl  & 72    & \nodata & \nodata & TS94  \\
$159.0-15.1$ & IC 351         & 1.31  & 2500   & 1.0    & 24.7 & 7 & wl  & 85    & \nodata & \nodata & TS94  \\
$194.2+02.5$ & J 900          & 1.13  & 1110   & 0.7    & 9.2  & 9 & wl  & 123   & \nodata & \nodata & TS94  \\
$221.3-12.3$ & IC 2165        & 1.41  & 3984   & 0.8    & 19.0 & 9 & wl  & 153   & \nodata & \nodata & TS94  \\
$356.2-04.4$ & Cn 2-1         & 0.97  & 5320   & 0.07   & 74.2 & 2.4 & wl  & 84    & \nodata & \nodata & P-M91 \\
$013.7-10.6$ & YC 2-32        & 0.88  & 3337   & 0.03   & \nodata & 15 & pn  & 68    & \nodata &  \nodata & TS94  \\
$037.7-34.5$ & NGC 7009       & 1.08  & 4371   & 0.3    & 151.8 & 28.5 & pn  & 85    & 2750 & $-8.55$ & TS94  \\
$084.9-03.4$ & NGC 7027       & 1.64  & 30000  & 1.1    & 2.8  & 14 & pn  & 175   & \nodata & \nodata & TS94  \\
$103.7+00.4$ & M 2-52         & 1.41  & 879    & 1.3    & 127.1 & 14 & pn  & \nodata & \nodata & \nodata &       \\
$104.4-01.6$ & M 2-53         & 1.11  & 496    & 0.1    & 1.6  & 14.8 & pn  & 112   & \nodata & \nodata & TS94  \\
$118.0-08.6$ & Vy 1-1         & 0.99  & 2101   & 0.0    & 56.6 & 5.2 & pn  & 32    & \nodata & \nodata & TS94  \\
$130.3-11.7$ & M 1-1          & 2.7:  & 3000   & 46.7   & 7.2  & 6 & pn  & 87    & \nodata & \nodata & TS94  \\
$133.1-08.6$ & M 1-2          & \nodata & 4264   & 0.02   & \nodata & 18 & pn  & 51    & \nodata & \nodata & TS94  \\
$147.4-02.3$ & M 1-4          & 1.03  & 6276   & 0.06   & \nodata & 4 & pn  & 67    & \nodata & \nodata & TS94  \\
$166.1+10.4$ & IC 2149        & 0.95  & 10000  & 0.0    & 6.8  & 8.5 & pn  & $<49$ & 1290 & $-7.7$ & TS94  \\
$196.6-10.9$ & NGC 2022       & 1.51  & 800    & 8.7    & 62.5 & 19 & pn  & 114   & \nodata & \nodata & TS94  \\
$215.2-24.2$ & IC 418         & 0.87  & 18000  & $>0.1$ & 0.44 & 12 & pn  & 53    & 1050 & $-8.2$ & TS94  \\
$243.8-37.1$ & PRTM 1         & 1.63  & 1200   & 7.8    & $>300$ & 23 & pn  & 90    & \nodata & \nodata & P90   \\
$294.1+43.6$ & NGC 4361       & 1.93  & 800    & $>19$  & $>180$ & 63 & pn  & 95    & \nodata & \nodata & TS94  \\
\bottomrule
\tabnotetext{a}{Nebular data are mostly from Paper II. Nebular diameter $\phi$ is from Acker et al. 1992. $T_\ast$, v$_\infty$ and log~$\dot M$ values were taken from literature,
and the references are: dM01: De Marco et~al. (2001); dMC99: De Marco \& Crowther (1999);  KH97a: Koesterke \& Hamann (1997a);
KH97b: Koesterke \& Hamann (1997b); L96: Leuenhagen et~al. (1996); L97a: Leuenhagen (1997a), L97b: Leuenhagen (1997b);
LH98: Leuenhagen \& Hamann (1998); P90: Pe\~na et~al. (1990); P-M91: Preite-Martinez et~al. (1991);
TS94: Tylenda \& Stasi\'nska (1994); ZK93: Zhang \& Kwok (1993).}
\tabnotetext{b}{Central star type is as following: a number indicates the [WC] spectral type, {\it wl} is for WELS
and {\it pn} for normal stars.}
\end{tabular}
}
\end{table*}

An ad-hoc detailed kinematical model, including the density distribution (3D-morphology),
and the ionization and thermal structures of the nebula, would be required to
deconvolve the effect of each contribution to the line profile. An approach to this procedure,
using a spherically symmetric model, can be found in Gesicki et al. (1996), Gesicki et al. (2003), and other papers by these authors. More recently, tomography
techniques or 3D spatio-kinematic models have been used to calculate the line profiles of some
individual nebulae (Sabbadin et~al.\@2004 and references therein; Monteiro et~al.\@ 2005).
The main weaknesses of such a procedure are that the results are strongly model dependent,
the derived model is probably not unique, and it is very much time consuming.

It is beyond the scope of this work to develop such a model for each of our objects.  
Our main aim is to derive reliable \Vexp \ for different ions in order to compare the behavior of WRPNe with those of non-WRPNe and to address, if
possible, the effects due to turbulence.

\smallskip

To relate the line profiles  with the values of the real expansion velocities for our 
objects in a consistent manner, we have used the code NEBU-3D, presented by Morisset et al.
(2005), to compute a spatio-kinematic model for analyzing  the line profile behavior in a 
simple and common case: an ellipsoidal nebula ionized by a star with an effective 
temperature of 80,000~K and a luminosity of 10$^{35}$ erg~s$^{-1}$. 
The chemical abundances are solar. The nebula has an internal elliptical hole with the 
major axis twice the minor axis. The orientation of the polar axis makes an angle of 
45$\degr$ with the line of sight (thus the nebula is in between ``polar-on'' and ``face on''
orientations). The density is constant in the radial direction, but varying with the polar 
angle according to N(H)$_{\rm inner}$ = N(H$_0$) $\times$ ($R_0$/$R_{inner})^2$, 
where N(H$_0$) and $R_0$ are the hydrogen density and the inner radius at the equator. 
The dependence of N(H) with the polar angle is included in the parameter $R_{inner}$ which 
follows the ellipsoidal form.

 Fig.~1 presents the H$\beta$ surface flux distribution, projected on the sky. The nebular 
dimensions projected on the sky are about $5''\times 7''$ (a distance of 1.87~kpc is assumed). 
Two ``observing slits'' (a $3''\times 4''$  slit, centered, and a similar one, off-center) have 
been superposed. These slits are used to extract ``observational'' data to compare with our own. 

The expansion velocity of the modeled nebula grows with the distance  from the star as 
V = 60 $R/R_{max}$ km~s$^{-1}$ ($R_{max}$ = 4.2 10$^{17}$~cm is the distance from the central 
star to the pole). A turbulent component of 0.5~km s$^{-1}$ has been added to the expansion.

We consider that this hypothetical nebula represents adequately the majority of nebulae. 
We have verified that for other morphologies (like spherical or bipolar nebulae) the 
results do not differ significantly from the conclusions presented here.  A very simple 
spherical case has been analyzed by Gesicki \& Zijlstra (2000) providing very similar results.
 An extensive catalog of profiles for elliptical and  bipolar objects seen in different orientations and 
with different conditions will be published elsewhere 
(Morisset et~al., in preparation).

The main characteristics of our modeled nebula are presented in Fig.~2. The upper panels 
show the density, temperature and velocity structures as function of the distance from the 
central star (in units of $R_{max}$) along the equator (solid lines) and along the polar 
axis (dotted lines). The second row of panels presents the surface brightnesses computed
for  H$\beta$, \ion{He}{II} $\lambda$4686, and [\ion{N}{II}] $\lambda$6583, as 
function of the distance from the central star along the same directions. 
The third row presents the line profiles of H$\beta$, \ion{He}{II}\,$\lambda$4686, 
and [\ion{N}{II}]~$\lambda$6583 as function of the velocity, obtained by integrating through
the two slits shown in Fig.~1. The solid lines present the profiles through the centered slit 
(this slit has the same size and position as our extraction window),  while the dotted lines 
show the profile through the off-center slit. The lowest row of panels shows the profiles 
through a $30''\times 30''$ slit which includes the whole nebula. Superposed to these profiles 
we have traced horizontal lines indicating the values of \Vexp \ as measured with different methods: the  
heavy solid line is the ``real velocity'' ---computed as the average velocity of the ion 
weighted by the emissivity of the line, the thin solid line is the  peak-to-peak velocity, the 
dotted line is  the HWHM velocity and the dashed line is the velocity at 1/10 the maximum 
intensity of the profile (\Vdiez, see \S~5).

As it is observed in Fig.~2, our model shows that when the nebula is
well centered in the slit and it is fully resolved (third row of panels) it presents symmetric
 double-peak profiles and that  half the peak-to peak separation is a very good approach for 
the ``real velocity'', underestimating it by less than 10\%. The HWHM of the profile overestimates 
the ``real velocity'' by about 10\%. The profiles obtained when the slit is 
off-center (dotted profiles) are always asymmetrical and it is very difficult to measure the 
expansion velocities from them, as the separation between peaks is smaller than the centered
 case. 

When the nebula is fully included in the slit, it presents single Gaussian profiles and 
HWHM of the profiles is a good measurement of \Vexp \ within 20\%.
As expected due to the adopted velocity law, [\ion{N}{II}]\,$\lambda$6583 presents a 
higher expansion velocity than \ion{He}{II}\,$\lambda$4686 and \Hbeta, and the latter one shows 
much wider lines due to the thermal broadening (which is included).

\Vdiez \ is always larger than the `real velocity' by about 30--40\%, but it is closer to 
the velocity of the most external (although faint) zones and it is useful to detect  
high-velocity low-emission components in the gas.


\begin{figure}[!t]
\includegraphics[width=\columnwidth]{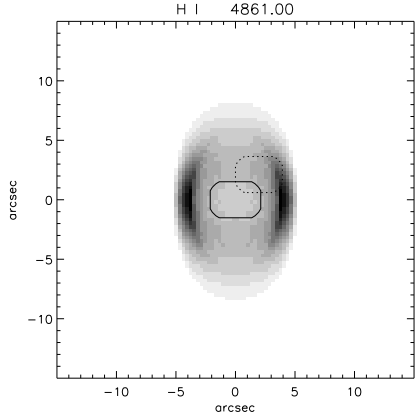}
\caption{The H$\beta$ surface brightness of the modeled nebula is shown. Two slits
of $3''\times 4''$ centered and off-center are superimposed. They are used to
produce ``observed profiles'' (see \S~3.1).}
\label{Fig.1}
\end{figure}

\begin{figure}[!t]
\includegraphics[width=\columnwidth]{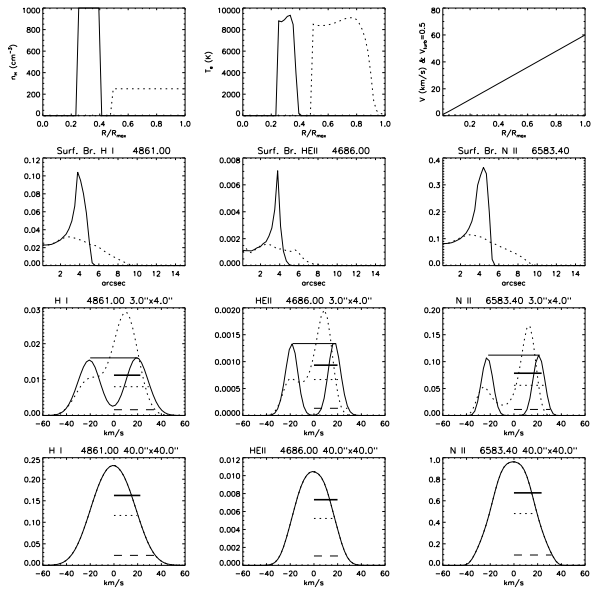}
\caption{Upper panels show the density, temperature and velocity distributions in the modeled nebula. The second row of panels shows the surface brightness along the equator (solid lines) 
and the polar axis (dotted lines) for several emission lines. The third row of panels shows 
the line profiles obtained through the centered slit (solid lines) and the off-center slit 
(dotted profile). The horizontal lines represent the ``real velocity'' (heavy solid line), the peak-to-peak (thin solid line), the HWHM (dotted line) and \Vdiez \ (dashed line) velocities, for the centered slit. 
The lowest panels show the line profiles through a slit of $30''\times 30''$, including the whole nebula.}
\label{Fig.2}
\end{figure}

\subsection{Our Observed Profiles}

Most of the profiles obtained for our objects fall into five different
categories:  a) symmetric double-peak profiles found mainly in extended objects, 
b) single Gaussian-type profiles usually found for compact symmetrical nebulae, the
prototype being M\,4-18 whose [\ion{N}{II}]$\lambda$6583 profile is shown in Fig.~3a, 
c) single asymmetrical profiles as seen in Fig.~3b, usually produced by asymmetric nebulae  
(or produced by aspherical nebulae where the slit did not cross exactly the center),
d) single profiles with wide wings or `shoulders' as shown in Fig.~3c (this could be due 
to high velocity gas or highly turbulent gas;  high velocity components have been reported 
for some objects, e.g., De Marco et al. 1997), e) complex
profiles showing several components which usually correspond to knotty nebulae
as the one presented in Fig.~3d.

For the first two categories, \Vexp\ can be measured safely in the classical form: 
from half the peak-to-peak separation for the well-resolved objects 
and from the HWHM  for objects showing single lines with no asymmetries nor complex 
profiles. Our measurements are presented in the next section. Objects with very asymmetrical 
profiles, for which it is very difficult to define \Vexp \ in a simple way, 
were excluded from the \Vexp\ analysis.
 For all the objects, the type of line profile is indicated in the last column of Table~5.

For the analysis, the profiles of \Hbeta, \ion{He}{II}\,$\lambda$4686, 
[\ion{O}{III}]\,$\lambda$5007, and [\ion{N}{II}]\,$\lambda$6583 lines were measured when 
available. The purpose for selecting these lines was to study possible
kinematical differences between lines arising from ions located at different distances from 
the central star.  For instance, while the \Hbeta\ line is mapping the whole nebula,
\ion{He}{II}\,$\lambda$4686 and [\ion{O}{III}]\,$\lambda$5007 lines give information about the 
inner and intermediate regions, and [\ion{N}{II}]\,$\lambda$6583, of the outer regions.

\section{Expansion velocities}

\subsection{ Objects with Split Profiles}
Twelve objects of our sample (six WRPNe, two WLPNe, and four ordinary PNe) present symmetrical 
split profiles.  According to our model such profiles are observed for well 
resolved nebulae and half the peak-to-peak separation is a good approach for \Vexp.
Values for [\ion{O}{III}]\,$\lambda$5007, H$\beta$, [\ion{N}{II}]\,$\lambda$6583, and 
\ion{He}{II}\,$\lambda$4686 lines are presented in Table~3. Typical errors are $\pm$5 \kms. 
The [WC] type of the central
stars or an indication if the star is a WELS or a ``normal'' central star is listed in Col.~3.

The line widths (FWHM) for each component of H$\beta$, $\Delta V_1$, and $\Delta V_2$ (in 
\kms), are also listed.  The instrumental and thermal widths have been subtracted from each
line width by assuming that they add in quadrature.  The instrumental widths were 
measured from the comparison lamp lines 
and they resulted to be 18 \kms \ for observations prior to 1997 (23$\mu$ pixel size CCD) 
and 12 \kms\ since 1997 (14$\mu$ pixel size CCD).  The thermal contribution to line width, for
any ion can be expressed as $\Delta V_{\mathrm{th}} = 21.4 \sqrt{t_{\mathrm{e}} / A}$~(\kms), 
where $t_{\mathrm{e}}$ 
is the electron temperature in units of 10$^4$\,K and \textit{A} is 
the atomic mass  (Lang 1980), therefore for H$\beta$,   
$\Delta V_{\mathrm{th}} = 21.4 \sqrt{t_{\mathrm{e}}}$ while for [\ion{O}{III}], 
$\Delta V_{\mathrm{th}} = 5.35 \sqrt{t_{\mathrm{e}}}$, etc.  The electron temperature employed 
for each object has been listed in Table~2. 

 The individual line widths in Table~3, in the range of 18--30 \kms, are
therefore mainly due to turbulence  and possible velocity gradients within the shell.   
In this sense notice that the model presented in \S~3, which has a velocity gradient producing  
$\Delta V \geq 20$ \kms\ from side to side of the shell, predicts a FWHM of about 5--6 \kms for the double-peak profiles
(third row of panels in Fig.~2), after subtracting the thermal width. Our objects in 
Table~3 present much higher $\Delta V$, possibly produced by turbulence.

Table 3 (see also \S~6.1) shows that, for a given object, the expansion velocities from 
different ions are, 
in general, very similar, although most of the objects have slightly larger
\Vexp([\ion{N}{II}]) than \Vexp([\ion{O}{III}]) or \Vexp(H$\beta$) which, as in our model, 
 might be indicating that the expansion velocity increases with the distance from the
central star.  This fact, predicted by hydrodynamical models as due to acceleration
of the external nebular material, was already noticed by Wilson (1950) for his
sample of PNe. 

Our \Vexp\ values are equal, within uncertainties with those from the catalogue
by Weinberger (1989), which have been included in Col.~10 of Table~3. In Col.~11 
we have also included the expansion velocities found by Acker et~al.\@  (2002),  based 
upon modeling the velocity field,  for the few objects in common; their \Vexp\ values 
and ours coincide, except for the faint and extended WRPN K\,2-16 for which Acker et~al. 
 give two different values. K\,2-16 is discussed in the Appendix~B.

For the sample in Table~3, we found that WRPNe have \Vexp\ values (as
measured from [\ion{O}{III}] lines) ranging from 24 to 44 \kms \ with an average of
36 \kms, while WLPNe and ordinary PNe show \Vexp\ from 17 to 26 \kms \ with an
average of 21.5 \kms.   That is $<$\Vexp(WRPNe)$>$ is 67\% larger than $<$\Vexp(non-WRPNe$>$).
 Therefore, definitely, WRPNe in this sample have larger expansion velocities 
than non-WRPNe (WLPNe and ordinary PNe).  Also a systematic trend of higher
\Vexp\ with earlier [WC] type seems to be present. This result will be strengthened in 
the following sections. 

In addition to higher \Vexp, WRPNe seem also to show larger line widths than non-WR
objects. This is particularly true for WRPNe with early [WC] stars such as NGC\,6905, 
NGC\,6369, NGC\,6751, and NGC\,1501, which in principle are more evolved, probably indicating large turbulence in these nebulae.

In conclusion, WRPNe with split profiles are showing higher \Vexp \ and 
probably more turbulence than WLPNe and ordinary PNe. This does not agree with the results 
presented by Acker et~al.  (2002), who have concluded that the expansion velocities in WRPNe and 
non-WRPNe are similar but WRPNe are much more turbulent than non-WRPNe.  
Also their 
 models would indicate that ordinary PNe present acceleration in the 
outer nebular zones, while WRPNe do not, which is not the case for our objects.

\begin{figure*}[!t]\centering
     \begin{tabular}{ll}
       \includegraphics[width=6cm,height=6cm]{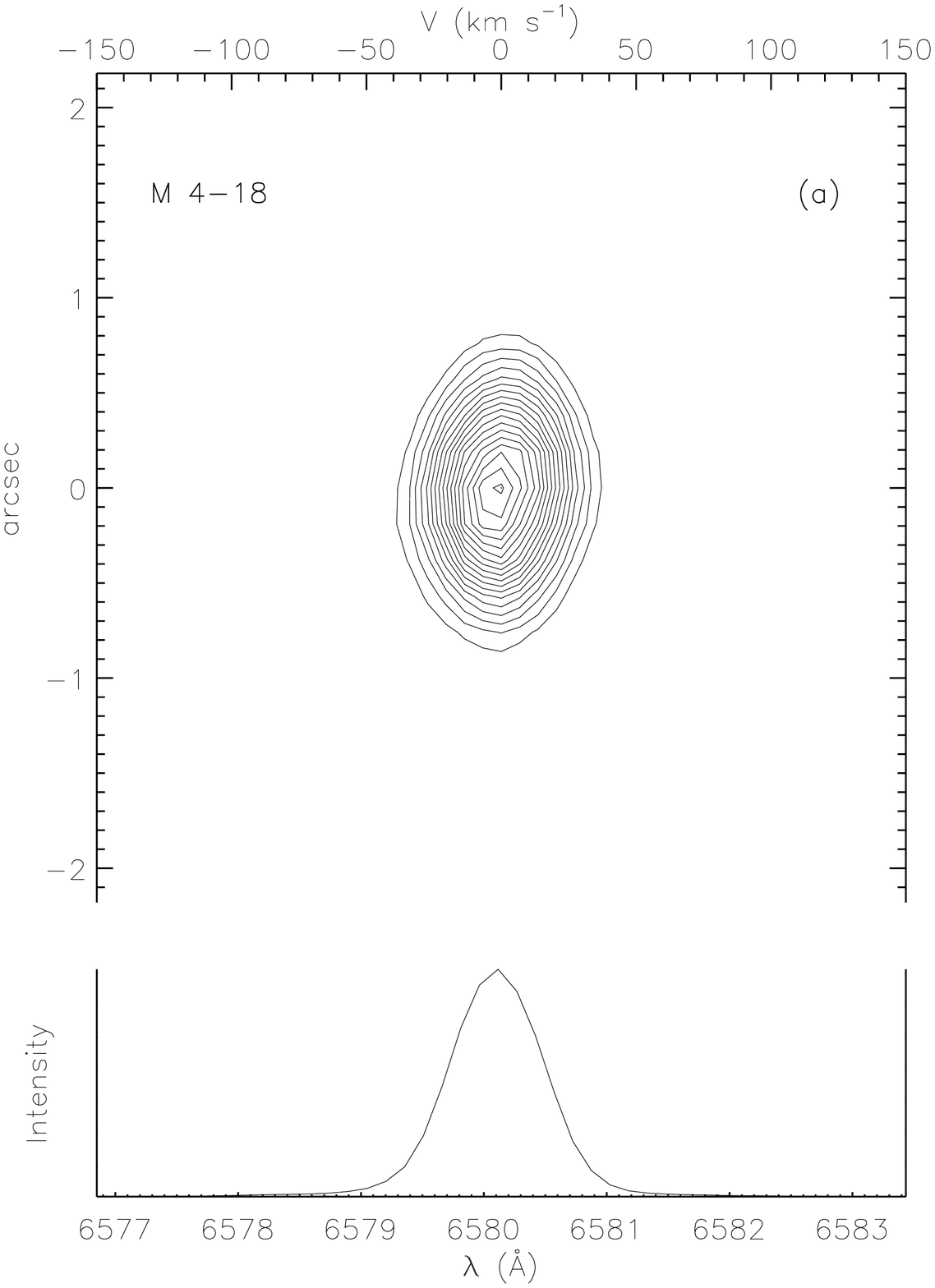} &
       \includegraphics[width=6cm,height=6cm]{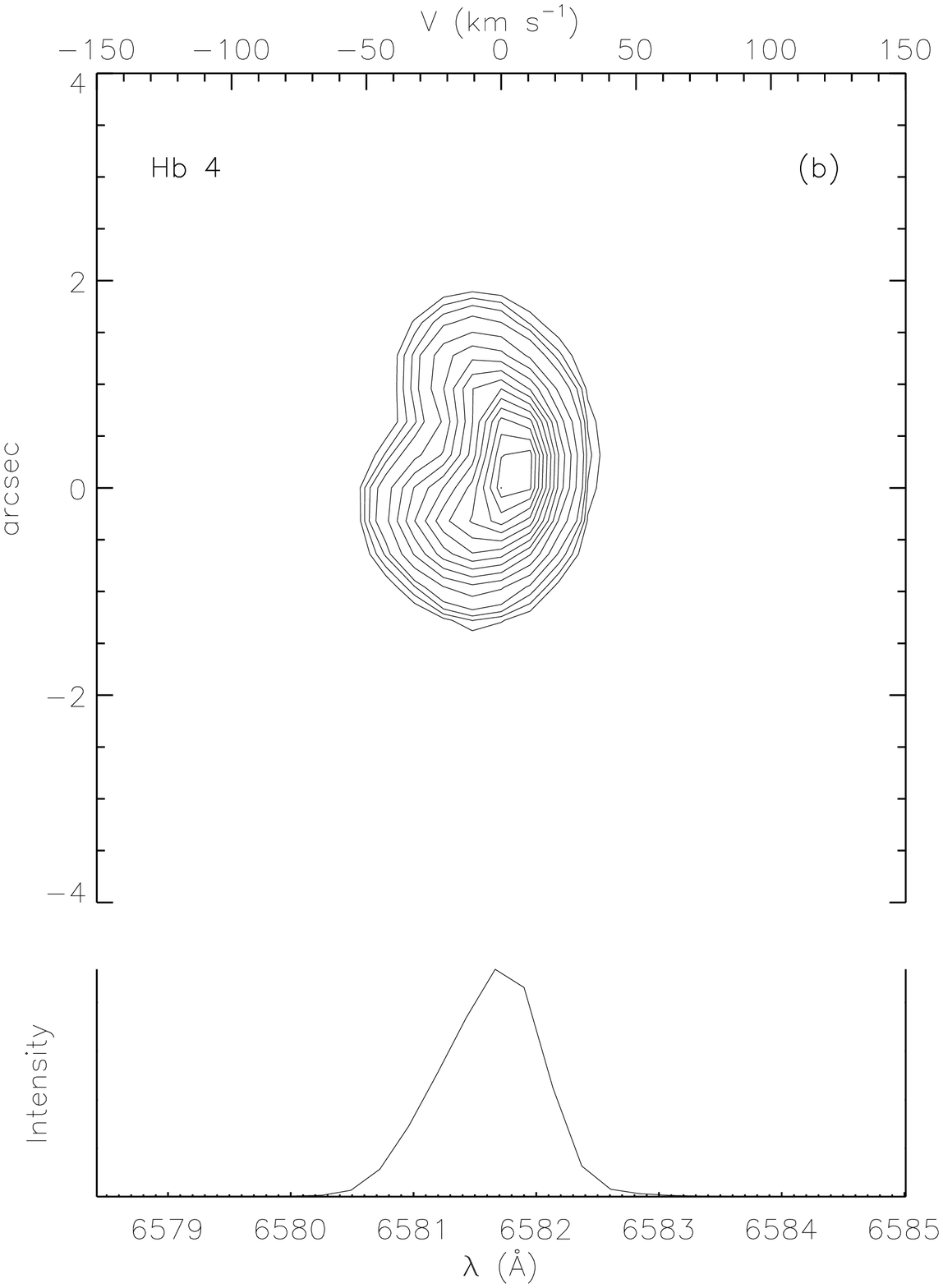} \\
       & \\
       \includegraphics[width=5cm,height=6cm]{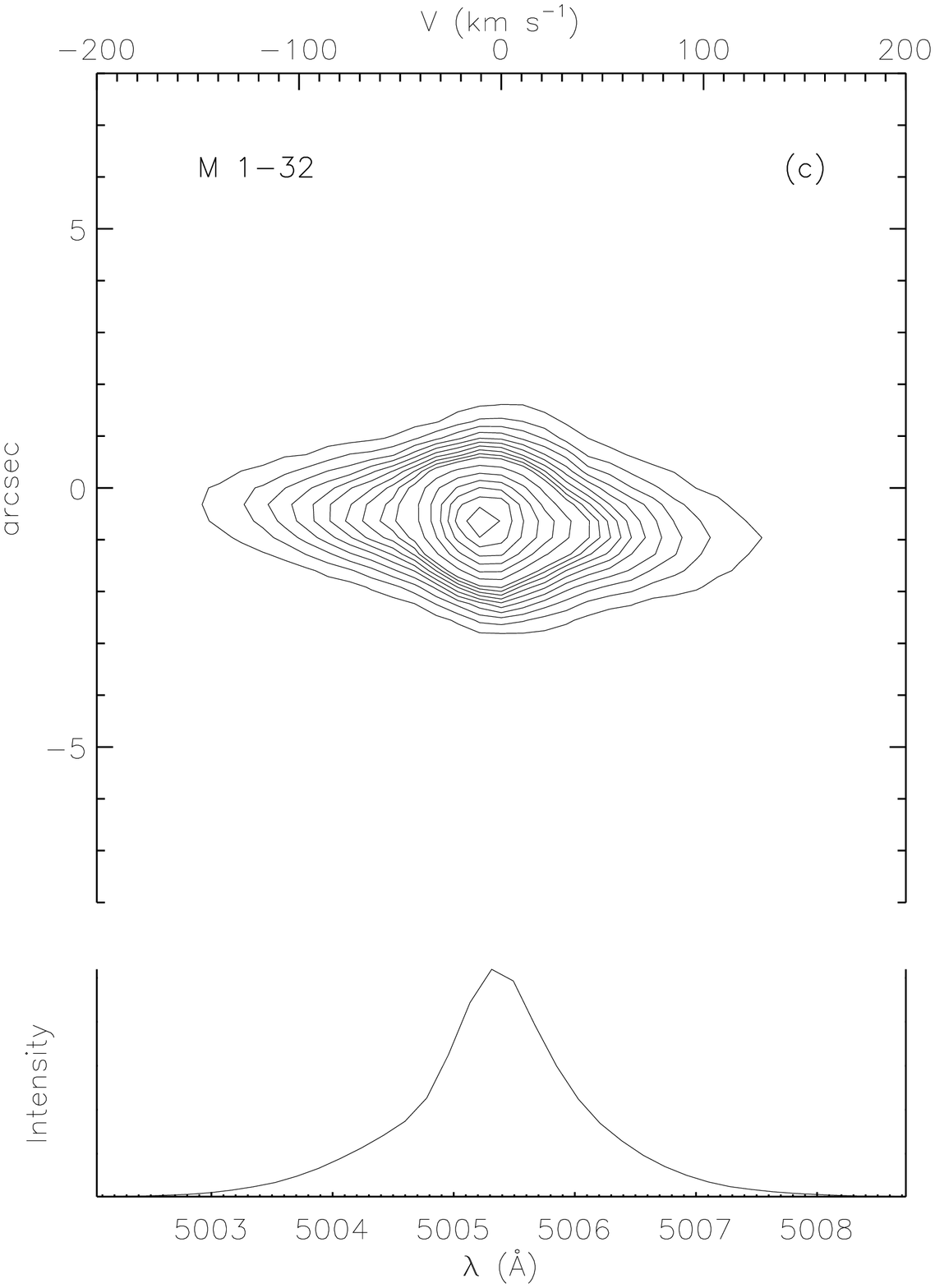} &
       \includegraphics[width=5cm,height=6cm]{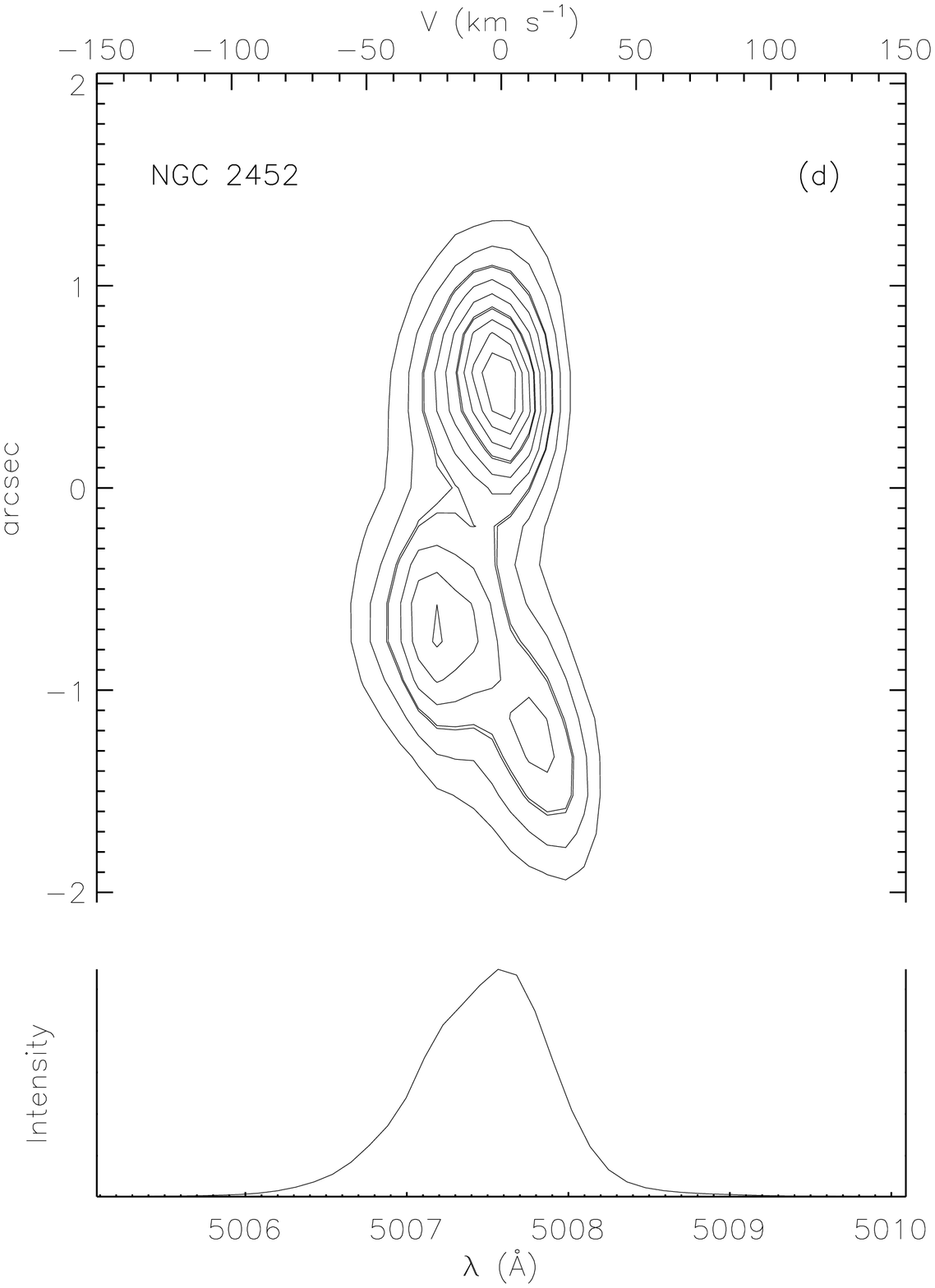} \\
       & \\
     \end{tabular}
   \caption[]{Position-velocity diagrams and integrated profiles from the extracted spectra are shown featuring the different types of line profiles
   found in our sample. The different types are:  a) simple Gaussian [{\ion{N}{II}}]\,$\lambda$6583 profile of M\,4-18, b) asymmetrical
   [{\ion{N}{II}}]\,$\lambda$6583
   profile of Hb\,4, c) [{\ion{O}{III}}]\,$\lambda$5007 high velocity wings found in
M\,1-32, d) knotty structure and complex
   [{\ion{O}{III}}]\,$\lambda$5007 profile in NGC\,2452. At the top of each figure, a velocity scale helps to visualize the velocity profile. The 0 corresponds to the maximum line intensity.}
\end{figure*}

\begin{table*}[!t]\centering
\setlength{\tabnotewidth}{0.8\textwidth}
\setlength{\tabcolsep}{0.5\tabcolsep}
\tablecols{11}
  \caption{Expansion velocities and turbulence  for nebulae with double peak lines}
\begin{tabular}{llccccccccc}
\toprule
PN G & Name & [WC]\tabnotemark{a} & \multicolumn{4}{c}{ \Vexp\,(\kms)\tabnotemark{b}} & $\Delta V_1$\tabnotemark{b} & $\Delta V_2$\tabnotemark{b}  & \Vexp(OIII)\tabnotemark{c} & \Vexp\tabnotemark{c}\\
   \cline{4-7}
     &     &     & [\ion{O}{III}] & H$\beta$ & [\ion{N}{II}] &  \ion{He}{II} &  H$\beta$ & H$\beta$ &W89 & A02 \\
\midrule
$061.4-09.5$& NGC\,6905   & 2-3 & 44& 41& 46& 40& 25 & 17 & 44 & \nodata \\
$002.4+05.8$& NGC\,6369   & 4   & 36& 36& 37& 38& 24 & 38 & 42 & \nodata   \\
$029.2-05.9$& NGC\,6751   & 4   & 42& 40& 40& \nodata & 24 & 29 & 40 & 41  \\
$144.5+06.5$& NGC\,1501   & 4   & 40& 40& 42& 38& 20 & 19 & 37 & 40  \\
$120.0+09.8$& NGC\,40     & 8   & 30& 26& 26& \nodata & 18 & $<15$ & 29 & \nodata  \\
$352.9+11.4$& K\,2-16     & 11  & 24& 24& 26& \nodata & 22 & 16 & \nodata &34 or 38 \\
$100.6-05.4$& IC\,5217    & wl        & 18& 18& 23& 18& 16 & 20 & 23 & \nodata \\
$011.7-00.6$& NGC\,6567   & wl        & 17& 18& 27& \nodata & $<20$ & $<20$ & 19 & \nodata \\
$037.7-34.5$& NGC\,7009   & pn        & 20& 18& 21& 16& $<15$ & $<15$ & 21 & \nodata \\
$196.6-10.9$& NGC\,2022   & pn        & 24& 21& 21& 23& 16 & 20 & 27 & \nodata \\
$243.8-37.1$& PRTM\,1     & pn        & 24& 23& \nodata & 23& 28 & 26 & \nodata & \nodata \\
$294.1+43.6$& NGC\,4361   & pn        & 26& 22& \nodata & 22& 32 & 20 & 21 & \nodata \\
\bottomrule
\tabnotetext{a}{Central star type as in Table 2.}
\tabnotetext{b}{Uncertainties: $\pm5$\,\kms \ for \Vexp, and $\pm8$\,\kms \ for line widths.
$\Delta V$s were corrected for thermal and instrumental widths.}
\tabnotetext{c}{References are: W89: Weinberger (1989); A02: Acker et~al. (2002).}
\end{tabular}
\end{table*}

\subsection{Objects with Single Profiles}

The FWHM of [\ion{O}{III}]\,$\lambda$5007, H$\beta$, [\ion{N}{II}]\,$\lambda$6583, and \ion{He}{II}\,$\lambda$4686 were measured for all the objects showing single 
symmetrical profiles  ([\ion{N}{II}]\,$\lambda$6548 was measured when 
the stellar \ion{C}{2} lines were severely blended with [\ion{N}{2}]$\lambda$6583. See Appendix~A). 
Instrumental and thermal widths were subtracted as indicated in \S~4.1.  
We have assumed that HWHM represents the expansion velocity plus turbulence.
The results are presented in Table~4 where, in Col.~8,  we have included \Vexp([\ion{O}{III}]) 
from the compilation by Weinberger (1989) or from more recent literature and in Col.~9
the values reported by Acker et~al.\@(2002).  

Although in this case it is more difficult to compare kinematical data measured by
different authors using different procedures, it is found that the majority of our values
coincides within a factor of 1.2 with those from the literature.  For several objects  
(M\,1-32, He\,2-459, K\,1-61, and YC\,2-32), \Vexp\ are  determined for the first time.

 Typical errors for our measurements are $\pm$6--8 \kms\ for  [\ion{N}{2}] and \ion{He}{2} 
lines and $\pm$4 \kms\ for [\ion{O}{3}] and \Hbeta\ lines. We have searched for systematic 
errors (as, e.g., a dependence of \Vexp\ as a function of electron temperatures) and found that these  errors are negligible. Also possible artificial broadening of nebular lines due to contamination with  stellar  lines, is discarded as a source of important systematic 
errors (see Appendix~A). 
Systematic trends of \Vexp\ with electron density and stellar temperature are found for all the sample. These  are  real trends and they are discussed in the next sections.

 \Vexp\ averages of the 14 WRPNe in Table~4 are 23, 23, and 25\,\kms, for [\ion{O}{III}], 
H$\beta$ and [\ion{N}{II}], respectively.  The corresponding averages for the 5 WLPNe are 19, 18,
 and 22 and for the ordinary PNe are: 19, 17, and 20. WLPNe and ordinary PNe show a very similar  behavior and can be put together in a single sample, non-WRPNe, showing averages of   
19, 18, and 20\,\kms\ for [\ion{O}{III}], H$\beta$, and [\ion{N}{II}]. Therefore,
 the averages for WRPNe are 20--25\% larger than the corresponding
averages for non-WRPNe indicating that WRPNe have wider lines.  In this case we cannot 
disentangle expansion velocities from turbulence, but this result is consistent with the 
results found in \S~4.1 of higher expansion velocities and probably higher turbulence in 
WRPNe than in non-WRPNe.

As in the case of double-peak objects, here the \Vexp\ average for [\ion{N}{II}] is 
about 2\,\kms \ larger than the corresponding values for [\ion{O}{III}], regardless 
if they are WRPNe or non-WRPNe, confirming that the external shells might be expanding faster 
in all kind of nebulae.

\section{Line widths at the base of lines}
Considering that an important fraction of PNe shows kinematical features like ansae,
BRETS, FLIERS, jets, high velocity ejecta, etc.,  (see Balick \& Adam 2002 for a 
definition of these terms), all of them produced by
non-spherical, bipolar, multipolar or asymmetrical ejections of the central star,
it seems likely to expect some additional perturbations in the velocity field of
WRPNe as compared with non-WRPNe, caused by the [WC] wind.  For instance,
the hydrodynamical models of Garc{\'\i}a-Segura \& McLow (1995) show
that the expansion of a hot bubble pushed by a WR wind results in a filamentary
broken shell with large turbulence.

Therefore, to perform a kinematical analysis of all our sample, 
including all these effects, we decided to consider the nebular gas showing the
highest velocity relative to the star. This can only be measured at the base of 
the line profiles but to avoid any noise disturbance
 due to the low signal at the base of the lines, we measured such a velocity as half the  
width of the line profile at one tenth maximum intensity (\Vdiez), after
subtracting the instrumental  and  thermal widths as described in \S~3.1.
A similar treatment was previously used by Dopita et~al.\@ (1985) to determine 
expansion velocities of a sample of PNe in the SMC.

\Vdiez \ was 
determined for all the objects, including those cases where a single Gaussian was
adequately reproducing the line profiles or when the lines showed
a double peak.  In the latter case, half the
full width at 1/10 intensity of each component was added to the peak-to-peak
separation. It can be easily calculated that, for a single Gaussian profile,
half the full width at 1/10~I max can be expressed as:
\begin{displaymath}
V_{10}   = 0.911 \ {\mathrm{FWHM}}\ \ ({\mathrm{km}}\,{\mathrm s}^{-1}) \, ;
\end{displaymath}
while for double peak profiles (two Gaussians), it is found that:
\begin{displaymath}
V_{10}  = V_{\mathrm{exp}} + 0.455 \left( {\Delta V_1} + {\Delta V_2} \right) \ \ ({\mathrm{km}}\,{\mathrm s}^{-1})\, .
\end{displaymath}
We have used these expressions to derive \Vdiez\ for our objects in Tables~3 and 4. 
For non-Gaussian, wing-extended, or complex profiles, \Vdiez \ was
measured directly from the line profiles.

\Vdiez \ derived in such a way includes not only \Vexp \  of the
shell, but also the turbulence of the gas and faint high velocity components if they exist.
The values are listed in Table~5.  Uncertainties are about $\pm$5~\kms \ and larger if
marked with a colon.  In the last column of Table~5 we describe the line profiles:
\underline{s} for single, \underline{d} for double peak, \underline{c} for complex,
\underline{h} for high velocity components. 

\begin{table*}[!t]\centering
\setlength{\tabnotewidth}{0.7\textwidth}
  \setlength{\tabcolsep}{0.5\tabcolsep}
  \tablecols{9}
\caption{Expansion velocities  in objects with single symmetrical lines}
\begin{tabular}{clcccccrr}
\toprule
PN G & Main Name & [WC]\tabnotemark{a} & \multicolumn{4}{c}{\Vexp\, (\kms)\tabnotemark{b}} &  \Vexp\tabnotemark{c} & \Vexp\tabnotemark{d} \\
  \cline{4-7}
& &  &  [\ion{O}{III}] & H$\beta$ & [\ion{N}{II}] &  \ion{He}{II} & [\ion{O}{III}] & \\
& &  & $\lambda$5007 & $\lambda$4861 & $\lambda$6583 &  $\lambda$4686 & W89 & A02 \\
\midrule
$001.5-06.7$ & SwSt 1  & 9   &  12   & 12 & 17   & \nodata & 13$^1$ & 17   \\
$002.2-09.4$ & Cn 1-5  & 4   &  23   & 22 & 24   & \nodata & 18:  & \nodata \\
$004.9+04.9$ & M 1-25  & 6   &  23   & 23 & 23   & \nodata  & \nodata  & 30  \\
$006.8+04.1$ & M 3-15  & 5   &  17   & 17 & 18   & \nodata  &   \nodata & 16  \\
$011.9+04.2$ & M 1-32  & 4-5 & $\leq13$ & $\leq13$ & $\leq13$ & \nodata  & \nodata & \nodata \\
$012.2+04.9$ & PM 1-188& 10  &  \nodata   & 30 & 38   & \nodata  &  \nodata & \nodata \\
$027.6+04.2$ & M 2-43  & 8   & 14    & 15 & 17   & \nodata  & \nodata & 20 \\
$048.7+01.9$ & He~2-429 & 4-5  & 28    & 29 & 32   & \nodata & \nodata & \nodata \\
$064.7+05.0$ & BD+30\degr3639 &9 & 43    & 23 & 23  & \nodata  & (e) & 28 \\
$068.3-02.7$ & He 2-459   & 8  & \nodata & 32 & 33   & \nodata  & \nodata & \nodata \\
$096.3+02.3$ & K 3-61     &4-5 & 28  & 32 & 25   &  \nodata  &  \nodata  & \nodata \\
$130.2+01.3$ & IC 1747    & 4  & 30  & 29 & 30   &  30  & 28~ & \nodata \\
$146.7+07.6$ & M 4-18     &10  & \nodata  & 12 & 13   & \nodata  & 19$^2$ & 15  \\
$010.8-01.8$ & NGC 6578   & wl  & 16  & 16    & 19 & \nodata   &  \nodata & \nodata  \\
$159.0-15.1$ & IC 351     & wl  & 20  & 14    & 16: & 19  &   15~ & \nodata \\
$194.2+02.5$ & J 900      & wl  & 23  & 24    & 27 & 22   &  13~ & \nodata \\
$221.3-12.3$ & IC 2165    & wl  & 24  & 24    & 26 & 24   &  20~ & \nodata \\
$356.2-04.4$ & Cn 2-1     & wl  & 14  & 14    & \nodata & 18:  & 23$^3$ & \nodata \\
$013.7-10.6$ & YC 2-32    & pn  & 18  & 18    & 22  & 20   & \nodata   & \nodata \\
$084.9-03.4$ & NGC 7027   & pn  & 22  & 21    & 28  & 22   & 22~ & \nodata \\
$103.7+00.4$ & M 2-52     & pn  & 19  & 19    & 16  & 22   & 8~ & \nodata \\
$104.4-01.6$ & M 2-53     & pn  & 17  & 21    & 20: & 22:  & 11~ & \nodata  \\
$118.0-08.6$ & Vy 1-1     & pn  & 12  & 13    & 18: & \nodata   & 10~ & \nodata \\
$130.3-11.7$ & M1-1       & pn  & 31  & 30    & 22: & \nodata   & 39~ & \nodata \\
$147.4-02.3$ & M 1-4      & pn  & 15  & 14    & \nodata & 11   &  14~ & \nodata  \\
$166.1+10.4$ & IC 2149    & pn  & $<9$~~~ & 11  & 18 & \nodata   & $<10$~ & \nodata   \\
$215.2-24.2$ & IC 418     & pn  & $<9$~~~ & 8  & 14 & \nodata  & $<~6$~  & 15 \\
\bottomrule
\tabnotetext{a}{Central star type as in Table 2.}
\tabnotetext{b}{Typical errors are $\pm$6--8 \kms\ for [\ion{N}{2}] and \ion{He}{2}, and $\pm$4 \kms\ for 
H$\beta$,  and [\ion{O}{3}]. Errors are larger for the cases marked with a colon.}
\tabnotetext{c}{Data are from Weinberger (1989) unless indicated otherwise. Other references are:
1: De Marco et~al. (2001); 2: De Marco \& Crowther (1999);    
3: Gesicki \& Zijlstra (2000).}
\tabnotetext{d}{A02: Acker et~al. (2002).}
\tabnotetext{e}{ Bryce \& Mellema (1999) reported \Vexp([\ion{O}{III}])=36, \Vexp([\ion{N}{II}])=28 \kms\ for BD+30\degr3639.} 
\end{tabular}
 \end{table*}

\bigskip

\begin{table*}[!t]\centering
\setlength{\tabnotewidth}{0.8\textwidth}
  \tablecols{10}
\caption{\Vdiez \ for the selected ions}
\footnotesize{
\begin{tabular}{cccccccccc}
\toprule
PN G & Usual Name & [WC]\tabnotemark{a} & \multicolumn{6}{c}{\Vdiez\,(\kms)\tabnotemark{b}} &  profiles\tabnotemark{c} \\
   \cline{4-9}
& &  & [\ion{O}{II}] & [\ion{O}{III}] & H$\beta$ & [\ion{N}{II}] & \ion{He}{I} & \ion{He}{II} &  \\
& &  & $\lambda$3729 & $\lambda$5007 & $\lambda$4861 & $\lambda$6583 & $\lambda$5876 & $\lambda$4686 &   \\
\midrule
\renewcommand\baselinestretch{1.17}\small
$001.5-06.7$ & SwSt 1         & 9   & 29  & 22    & 22 & 30   & \nodata  & \nodata  & s   \\
$002.4+05.8$ & NGC 6369       & 4   & 76: & 64    & 64 & 57   & 67  & 63: & d,c  \\
$002.2-09.4$ & Cn 1-5         & 4   & 46  & 42    & 41 & 44   & \nodata  & \nodata  & s   \\
$003.1+02.9$ & Hb 4           & 3-4 & 47: & 32    & 32 & 34   & 31  & 42  & s,a  \\
$004.9+04.9$ & M 1-25         & 6   & 36  & 41    & 41 & 41   & 43  & \nodata  & s \\
$006.8+04.1$ & M 3-15         & 5   & \nodata  & 30    & 32 & 29   & 36  & \nodata  & s,h  \\
$011.9+04.2$ & M 1-32         & 4-5 & 63  & 89    & 62 & 65   & 76  & \nodata  & s,h  \\
$012.2+04.9$ & PM 1-188       & 10  & \nodata  & \nodata    & 56 & 70   & \nodata  & \nodata  & s   \\
$017.9-04.8$ & M 3-30         & 2  & 41: & 52    & 49 & 68   & 49  & 52  & c  \\
$027.6+04.2$ & M 2-43         & 8   & 26  & 26    & 27 & 30   & 30  & \nodata  & s   \\
$029.2-05.9$ & NGC 6751       & 4   & 64  & 66    & 63 & 60:  & 81: & \nodata  & d   \\
$048.7+01.9$ & He~2-429       & 4-5 & \nodata  & 52    & 54 & 59   & \nodata  & \nodata  & s   \\
$061.4-09.5$ & NGC 6905       & 2-3 & 64  & 64    & 60 & 64   & 68  & 67  & d   \\
$064.7+05.0$ & BD+30\degr3639 & 9   & 62  & 79    & 43 & 42   & \nodata  & \nodata  & s  \\
$068.3-02.7$ & He 2-459       & 8   & \nodata  & \nodata    & 56 & 60   & \nodata  & \nodata  & s   \\
$089.0+00.3$ & NGC 7026       & 3   & 68  & 54    & 55 & 62   & 59  & 52  & d,c \\
$096.3+02.3$ & K 3-61         & 4-5 & \nodata  & 48    & 57 & 46   & 51  & \nodata  & s   \\
$120.0+09.8$ & NGC 40         & 8   & 37: & 47    & 38 & 36   & 40  & \nodata  & d   \\
$130.2+01.3$ & IC 1747        & 4   & 25: & 55    & 54 & 58   & 55  & 56  & s,a \\
$144.5+06.5$ & NGC 1501       & 4   & \nodata  & 58    & 54 & 51   & 61  & 66  & d   \\
$146.7+07.6$ & M 4-18         & 10  & 24  & \nodata    & 21 & 24   & \nodata  & \nodata  & s   \\
$161.2-14.8$ & IC 2003        & 3   & 40  & 42    & 39 & 44   & \nodata  & \nodata  & c   \\
$243.3-01.0$ & NGC 2452       & 2   & 40: & 56    & 58 & 55:  & 57  & 60: & c   \\
$352.9+11.4$ & K 2-16         & 11  & 50: & 44    & 42 & 48   & \nodata  & \nodata  & d   \\
$009.4-05.0$ & NGC 6629       & wl  & 34  & 25    & 30 & \nodata   & 31  & \nodata  & c   \\
$010.8-01.8$ & NGC 6578       & wl  & 32  & 29    & 30 & 34   & \nodata  & \nodata  & s   \\
$011.7-00.6$ & NGC 6567       & wl  & 38  & 35    & 34 & \nodata   & \nodata  & \nodata  & d   \\
$096.4+29.9$ & NGC 6543       & wl  & 43  & 34    & 33 & 43   & 33  & \nodata  & c   \\
$100.6-05.4$ & IC 5217        & wl  & 62: & 34    & 34 & 62   & 34  & 25  & d   \\
$159.0-15.1$ & IC 351         & wl  & 35  & 36    & 26 & 28:  & 34  & 34  & s   \\
$194.2+02.5$ & J 900          & wl  & \nodata  & 42    & 43 & 49   & 48  & 39  & s   \\
$221.3-12.3$ & IC 2165        & wl  & 47  & 44    & 43 & 47   & 44  & 43  & s   \\
$356.2-04.4$ & Cn 2-1         & wl  & 28  & 26    & 25 & 34:  & 29: & 32: & s   \\
$013.7-10.6$ & YC 2-32        & pn  & \nodata  & 33    & 32 & 40   & 34  & 36  & s  \\
$037.7-34.5$ & NGC 7009       & pn  & 36  & 33    & 30 & 38   & 33  & 29  & d   \\
$084.9-03.4$ & NGC 7027       & pn  & 53  & 40    & 38 & 51   & 44  & 40  & s   \\
$103.7+00.4$ & M 2-52         & pn  & 32  & 35    & 35 & 32   & 33  & 39  & s  \\
$104.4-01.6$ & M 2-53         & pn  & 36  & 30    & 37 & 36:  & 39  & 39: & s,a \\
$118.0-08.6$ & Vy 1-1         & pn  & 36: & 21    & 24 & 32:  & 26  & \nodata  & s   \\
$130.3-11.7$ & M 1-1          & pn  & 36: & 57    & 54 & 38:  & 57  & \nodata  & s   \\
$133.1-08.6$ & M 1-2          & pn  & \nodata  & 48    & 42 & 57:: & 41  & 49  & c   \\
$147.4-02.3$ & M 1-4          & pn  & \nodata  & 27    & 25 & 54:  & 34  & 19  & c  \\
$166.1+10.4$ & IC 2149        & pn  & 32  & $<18$~~~ & 20 & 33   & 26  & \nodata  & c   \\
$196.6-10.9$ & NGC 2022       & pn  & \nodata  & 40    & 37 & 38   & 47  & 49  & d   \\
$215.2-24.2$ & IC 418         & pn  & 19  & $<18$~~~ & 15 & 25   & 21  & \nodata  & s   \\
$243.8-37.1$ & PRTM 1         & pn  & \nodata  & 48    & 50 & \nodata   & \nodata  & 47  & d  \\
$294.1+43.6$ & NGC 4361       & pn  & \nodata  & 50    & 46 & \nodata   & \nodata  & 52  & d   \\
\bottomrule
\tabnotetext{a}{Central star type  as in Table 2.}
\tabnotetext{b}{Uncertainties of \Vdiez \ values are in average about of 4-5\,\kms, and larger for the cases marked with a colon.}
\tabnotetext{c}{Line profiles are as following: \underline{s}: single compact, \underline{a}: asymmetrical, \underline{h}: high velocity extensions,  \underline{d}: double peak, and \underline{c}: complex.}
\end{tabular}
}
\end{table*}

\section{Kinematical analysis}
Many authors have searched for correlations among the nebular
velocity field and the nebular and stellar parameters.  The reported results are not
always consistent.  For instance, Sabbadin (1984) and others reported a
correlation between the nebular radius and \Vexp, which has been interpreted as
changes in \Vexp\ as a consequence of nebular evolution.  Gesicki \&
Zijlstra (2000) have not found such a correlation for a sample of bulge PNe with well
determined distances.  Also acceleration of the external 
shells (appearing as larger \Vexp\ for  N$^+$ and
O$^+$, than for O$^{++}$) seems to be a well established fact for the majority of PNe.  
Nevertheless, from their kinematical model, Acker et~al.\@ (2002) reported no evidence of 
such an acceleration for WRPNe, but recently Gesicki et~al.\@ (2003) ---using a similar 
model for a sample of 14 PNe--- found that the line profiles of most nebulae in their 
sample can be better reproduced by adopting an U-shape velocity field,
 whether the central star is a [WC] or not.

Here we use our consistent data to  study the behaviour of \Vexp \ vs.
several properties of the nebulae and their central stars, searching for the physical 
causes affecting the velocity field in WRPNe.

\subsection{\Vexp \ from Different Ions}


\begin{figure}[!t]
\includegraphics[width=\textwidth]{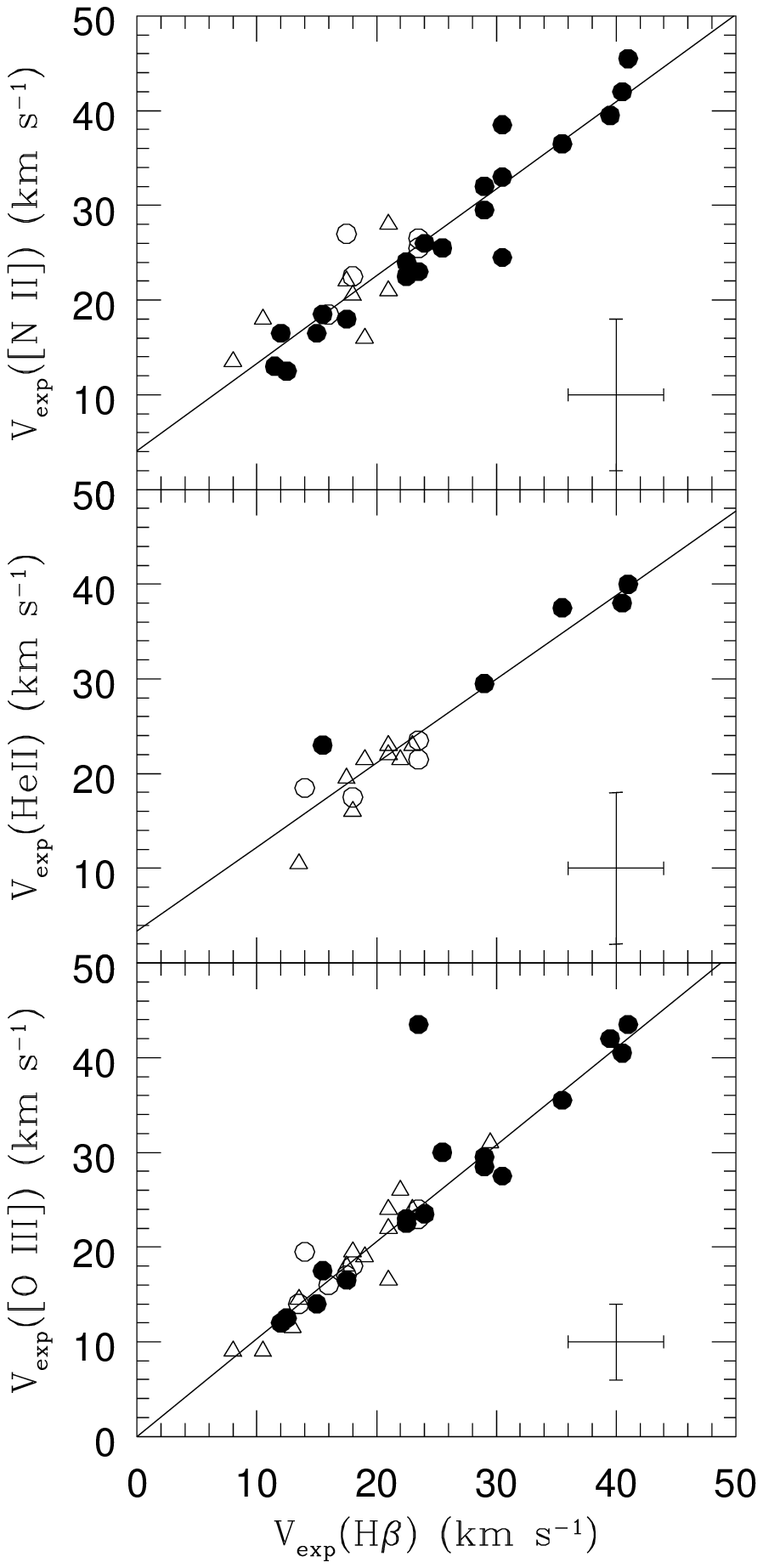}
\caption{\Vexp\  from \Hbeta\ are compared to \Vexp\  from:
[\ion{N}{II}], \ion{He}{II}, and [\ion{O}{III}]. WRPNe are marked with black circles, WLPNe are marked with open circles and ordinary PNe are marked with open triangles. Least squares fits are shown as solid lines. Typical error bars are shown in each case. The WRPN conspicuously out of the \Vexp([\ion{O}{3}]) versus \Vexp(\Hbeta) relation is BD+30\degr3639, discussed in Appendix~B.
}
\label{Fig.4}
\end{figure}

In Fig. 4, the values of \Vexp \ derived from [\ion{N}{II}], \ion{He}{II}, and [\ion{O}{III}] lines are plotted against the value derived from \Hbeta.  Different 
kind of objects (WRPNe, WLPNe, and PNe) are shown with different symbols  and typical error bars are presented in each graph ($\pm$8~\kms\ for [\ion{N}{II}] and \ion{He}{II}, and $\pm$4~\kms\ for [\ion{O}{III}] and \Hbeta).
  An immediate result from this figure is that \Vexp \ for WRPNe extends over a 
wide range from 10\,\kms \ up to 50~\kms, while \Vexp \ for ordinary PNe and WLPNe are 
always lower than 30~\kms. For the whole sample, the averages of \Vexp \ for WRPNe are 
about 40--45\% larger than the averages found for non-WRPNe.  Notice that, as already said,  $\langle$\Vexp(WRPNe)$\rangle$ / $\langle$ \Vexp(non-WRPNe)$\rangle $ is larger for 
double-peak objects 
than for  objects with single profiles. As double peak nebulae are more extended
 and presumably older objects,  this would indicate that evolved WRPNe expand faster
 that evolved non-WRPNe. 

Fig.~4 shows that \Vexp \ for [\ion{N}{2}], \ion{He}{2}, and \ion{O}{3}
follow a very close 45\degr~ linear correlation with \Vexp(\Hbeta),
although in the case of [\ion{O}{3}] versus  \Hbeta\ there is one WRPN, BD+30\degr3639,
that conspicuously departs from this behavior (see more details in the Appendix~B).  Interestingly, we do not find
significant differences among WRPNe, WLPNe, and PNe; all kind of objects present
the same correlations within uncertainties. The least square fits shown in Fig.~4 are:
\begin{itemize}
\item[$\bullet$] {$V_{\mathrm{exp}}({\mathrm{[\ion{N}{II}]}}) \   =  (0.92\pm0.08) \ V_{\mathrm{exp}}({\mathrm{H}}\beta) \ + \ (4.0\pm2.0)\, , \\ r=0.94$ \, ,}
\item[$\bullet$] {$V_{\mathrm{exp}}({\mathrm{\ion{He}{II}}}) \  =  (0.89\pm0.12) \ V_{\mathrm{exp}}({\mathrm{H}}\beta) \ + \ (3.0\pm2.9)\, , \\ r=0.95$ \, ,}
\item[$\bullet$] {$V_{\mathrm{exp}}({\mathrm{[\ion{O}{III}]}}) =  (1.02\pm0.08)\, V_{\mathrm{exp}}({\mathrm{H}}\beta) \ + \ (0.0\pm2.0)\, , \\ r=0.98$ \, .}
\end{itemize}
 In Fig. 4 and from the least square fits it can be seen that the [\ion{N}{2}] vs. \Hbeta\  
graph shows a larger dispersion but, within the errors, we find that in average, \Vexp([\ion{N}{2}]) is slightly larger than \Vexp(\Hbeta). [\ion{He}{2}] vs. \Hbeta\ correlation also might indicate that $\langle$\Vexp(\ion{He}{2})$\rangle$ is slightly larger than $\langle$\Vexp(\Hbeta)$\rangle$  but the uncertainties are larger in this graph and the result is not conclusive. On the other hand,
 \Vexp([\ion{O}{3}]) is equal, within uncertainties, to \Vexp(\Hbeta).  As already
 said, higher \Vexp\ for low ionization species were found in other PN samples
 (Wilson 1950; Gesicki et~al.\@ 2003), and could be due to acceleration of the  external 
zones relative to the inner shells or to higher turbulence of the external zones 
due to shocks or expansion of the ionization front against the neutral external material.  
We found that such a phenomenon is occurring in WRPNe as well as in non-WRPNe.  

The same statistical results are obtained if we use \Vdiez \ as given in Table~5. \Vdiez \ 
averages for WRPNe are  also about 40--45\% larger than the respective averages of non-WRPNe. 
As \Vdiez\ is computed for a larger number of objects the results are strengthened. 
Nevertheless the main importance in the analysis of \Vdiez \ compared to \Vexp \ is the 
possible detection of very extended wings or high velocity components produced by low emission 
gas, which are  only noticeable when \Vdiez \ is employed.
 This is discussed in more detail in the next sections.

\subsection{ Correlation of \Vexp \ with the Nebular Density}
In Fig.~5 (top), electron densities derived from the [\ion{S}{II}] line ratio are
plotted vs. \Vexp \ from \Hbeta.  The graph is largely scattered.  A wide range of
velocities corresponds to a given density, but it is evident that, regardless of the
stellar type (WR or not) the nebular density and \Vexp\ are anti-correlated.
As high density PNe are considered young nebulae, the
behavior found in Fig.~5 (top) is confirming an evolutionary effect in the sense that
low density (therefore evolved) nebulae expand faster and possibly have larger turbulence than
high density (young) objects.

In this figure, young high-density  WRPNe (all of them ionized by [WC]-late stars), present 
a wide range in velocities (from 12 to 30~\kms) while most young ordinary PNe and WLPNe are
constrained to \Vexp\,$\leq 20$~\kms.  Certainly, the powerful [WC] wind seems to largely affect
the velocity fields in seemingly young PNe. 
For the objects lying in the low density zone ($\log$ n$_{\rm e} \leq 3.5$), WRPNe appear 
much faster than normal PNe. Our sample includes only three WLPNe in this zone, but they are
showing the same behavior as ordinary PNe.

Fig. 5 (bottom) shows the behavior of density against \Vdiez(\Hbeta).
The velocity ranges are wider in this case, with \Vdiez\ from 20 to 65\,\kms \ for WRPNe,  while all WLPNe and PNe, except two, show values of $V_{10} \leq 45$\,\kms.
In both figures, the upper open triangle at log $n_{\mathrm{e}} = 4.48$,
corresponds to the young PN NGC\,7027 which shows an extraordinarily large \Vexp\
for its density.  It could be indicating the existence of a dense stellar wind.
This was also proposed by Keyes \& Aller (1990) in order to obtain an enhanced
stellar radiation field shortward 130 \AA, necessary to fit a photoionization model
for the nebula.  In recent observations of the central star, Hubeny et al. (1994)
did not detect such a wind, however its presence in the near past cannot be
discarded.

\begin{figure}[!t]
 \begin{tabular}{l}
\includegraphics[width=\columnwidth]{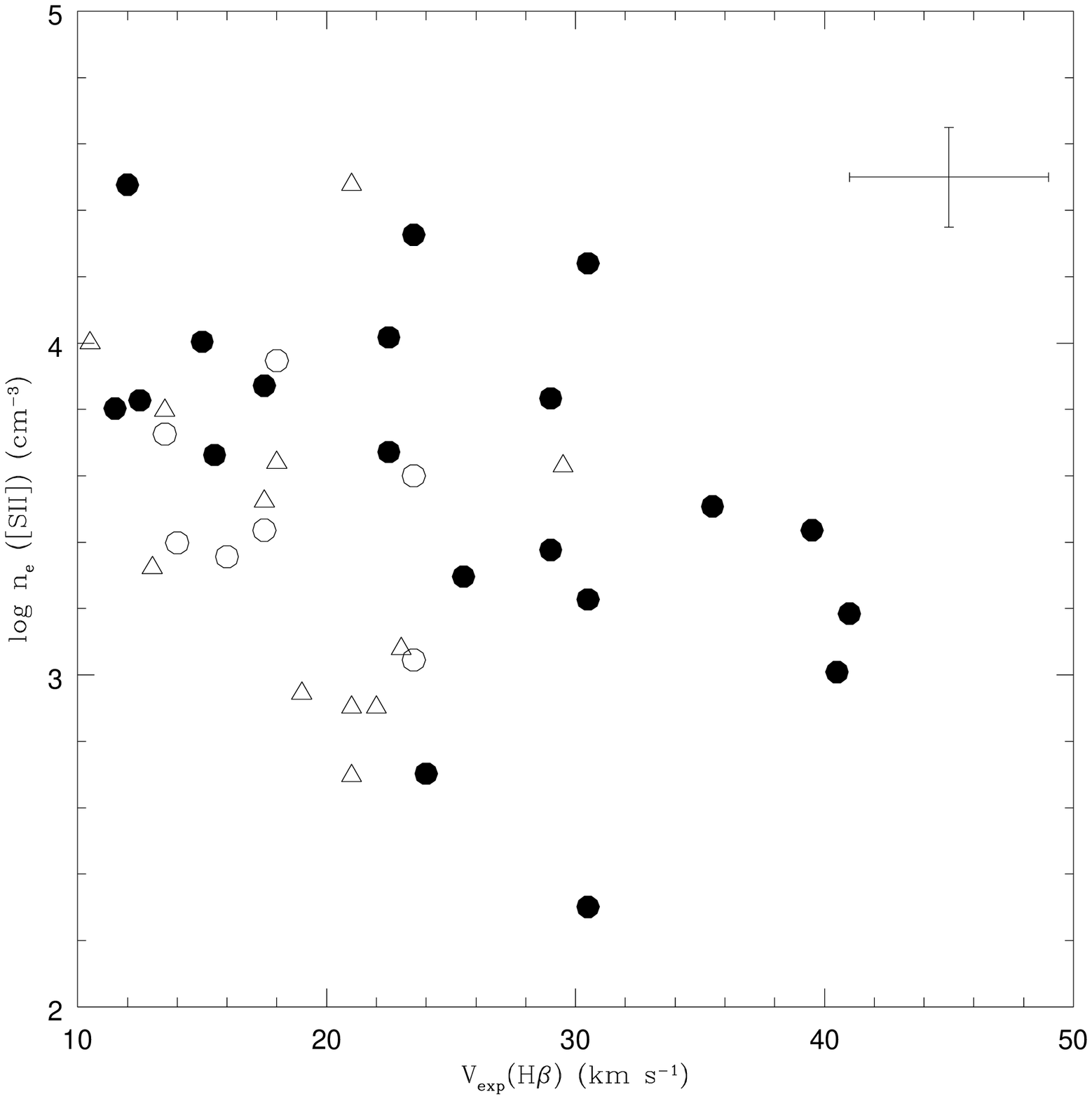} \\
\includegraphics[width=\columnwidth]{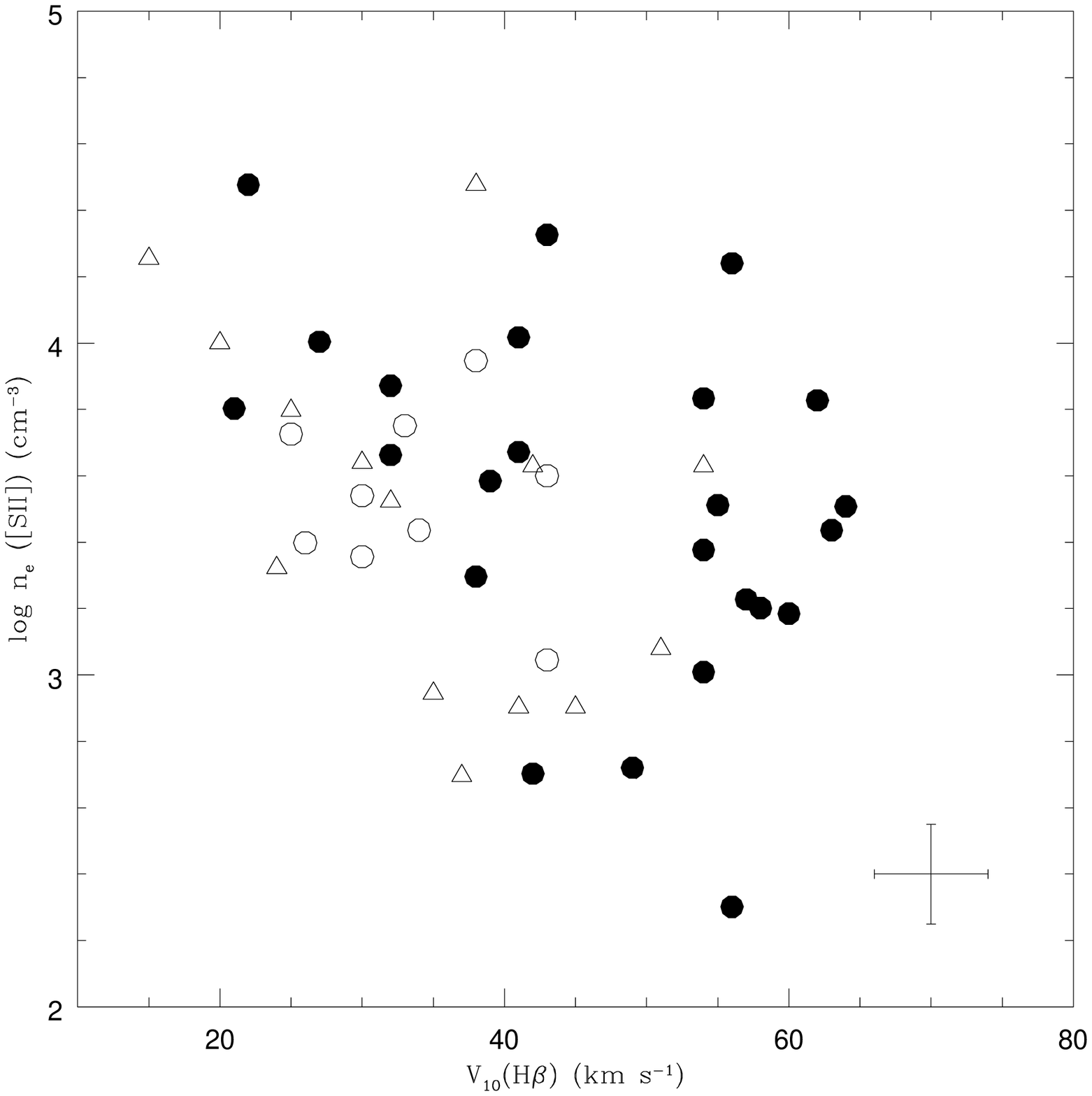}\\
\end{tabular}
\caption{Electron densities derived from [\ion{S}{II}] line ratio are plotted 
versus \Vexp\ (top) and \Vdiez\ (bottom)  from \Hbeta.
Symbols are as described in Fig.~4. Despite the large scatter, both quantities are anticorrelated. The upper open triangle at log~n$_e$=4.48 in both panels, corresponds 
to the young PN NGC\,7027.}
\label{Fig.5}
\end{figure}

\subsection{ Correlations of \Vexp\  with Characteristics of the
Central Stars}

\subsubsection{Stellar Temperature and Nebular Expansion}

The fundamental stellar parameters of many [WC] stars have been calculated from
non-LTE expanding atmosphere models by several authors (see references in Table~2).
It is found that stellar temperatures, $T_\ast$, are closely related to [WC] types
(e.g., \ Koesterke 2001).  It should be noticed that $T_\ast$ given by
models corresponds to the temperature at the base of the dense wind (optical
depth $\tau \sim 20$, e.g.,  Koesterke \& Hamann 1997b) and these are not
necessarily the temperatures of the ionizing fluxes ``seen'' by the nebulae.  
In fact there are some well documented [WC] stars
where the temperature at $\tau=2/3$ is much lower than $T_\ast$ and,
as a result, the PNe show  very low excitation.  A typical
case is NGC\,6751 which does not present nebular He$^{++}$ although it is ionized by a
[WC\,4] star of $T_\ast$=135,000\,K at $\tau=20$. Its $T_\ast$ at $\tau=2/3$ is, however, as
low as 55,500\,K (Koesterke \& Hamann 1997b).  This is very different from what occurs
in ordinary optically-thick non-WRPNe, where the \ion{He}{II} Zanstra temperature
and the stellar effective temperature are similar (Tylenda \& Stasi\'nska 1994).

To study the behaviour of \Vexp\ and \Vdiez \ as a function of $T_\ast$, we 
gathered the values of $T_\ast$ from the literature (see Table~2).
For [WC] stars which have been modeled, we adopted the model values.  For
the non-modeled [WC] stars and the other objects we have listed the \ion{He}{II} Zanstra
temperatures from the same data base as computed by Tylenda \& Stasi\'nska (1994).
Although this is a good approximation for the effective temperature for non-WRPNe,
it could be only a lower limit for the stellar temperature of [WC] stars.  

Fig.~6 presents the behaviour of \Vexp\ versus  $T_\ast$ (top) and \Vdiez\ versus $T_\ast$ (bottom).  
In spite of the uncertainties affecting $T_\ast$, a clear tendency
is noticed in both graphs showing that hotter stars ionize nebulae with
larger velocities and there is no object with a large $T_\ast$ showing \Vexp~$\leq$~20~\kms.
The scatter is however very large and the correlation coefficient for these graphs are
0.64 for \Vexp\  versus $T_\ast$ and 0.55 for \Vdiez\ versus $T_\ast$.

All nebulae, regardless if they are WRPNe or not, display this behaviour.
Two main groups can be distinguished in the graphs:  In the first, with
$T_\ast$ below 100,000\,K, some WRPNe show velocities as low as non-WRPNe and both 
type of objects appear mixed in the lowest velocity zone; in the the second group, with 
$T_\ast\,\geq$ 100,000\,K, WRPNe and non-WRPNe are well separated and WRPNe fall on a 
zone of much higher velocities than non-WRPNe.

Assuming that the stellar temperature is an indicator of the age of the
central star (which is the case for central stars evolving at approximately
constant luminosity), it is clear that the older the star, the faster
the nebular expansion, indicating an acceleration of the nebula with age for all type 
of stars. However for non-WR objects  the curve in Fig.~6 seems to flatten 
when the stars reach $T_\ast\,\sim$ 100,000 K. That is, non-WRPNe would stop accelerating 
and their \Vexp\  reach an upper limit of about 25--30 \kms\ while 
the acceleration continues for WRPNe. Evolved WRPNe show \Vexp\ of about 40 \kms.

This apparently continuous acceleration from young to evolved WRPNe is what one 
might expect if [WC] stars evolve from late to early types. A detailed hydrodynamical 
evolutionary model, including the kinematical effects 
of a long-term [WC] wind, should be compared with the behavior 
detected in Fig.~6, allowing us to confirm or discard the proposed evolutionary trend.  
Such a model is in progress (Medina et~al., in preparation) and will be published elsewhere.

The larger dispersion found in the \Vdiez\ versus $T_\ast$ graph shows that some objects, 
in particular some young WRPNe  and also a few ordinary PNe with relatively cold stars, 
present large turbulence and/or high velocity components since early phases of evolution. 
These phenomena are present in many PNe  irrespective of their age.

Acker et~al.\@  (2002) have not found any correlation between \Vexp\ and $T_\ast$ 
in their data, because they have used the stellar temperatures obtained
from photoionization models.  That is, they used $T_\ast$ ``seen'' by the nebulae,
which, as explained above, could be several thousands degrees lower than the
temperature at the base of the wind of the [WC] star.  Any possible relation is 
thus destroyed.

\subsubsection{[WC] Mass-Loss Rate and Wind Velocity versus Nebular Characteristics}

It is worth to notice that the kinetic energy carried by the particles in the
stellar winds can easily ionize the innermost gaseous material thus perturbing the
ionization equilibrium.  Therefore, to analyze this possibility, in Fig.~7 we present 
the nebular ionization degree He$^{++}$/He$^+$ of WRPNe as a function of the mass-loss rate,
$\dot M $, and the terminal velocity of the wind, v$_\infty$ (these parameters
are listed in Table~2).  It is evident in these graphs that highly ionized WRPNe
are associated with stars with low mass-loss rates ($\log$ $\dot M \leq -6$) and
v$_\infty$ larger than 1800\,\kms.  This is expected if we consider that such wind
parameters are typical of [WC]-early (therefore hot) stars, which produce
very excited nebulae with large \Vexp.  On the other hand, [WC]-late stars which have
higher $\dot M$ and lower v$_\infty$ only produce low ionization PNe
with no traces of He$^{++}$. Therefore, we have found that the powerful stellar winds 
do not contribute significantly to the ionization of nebulae and the ionization can be safely 
attributed to the UV stellar photons.

 \subsection{The Mechanical Energy of the Stellar Wind \linebreak
             and the Expansion Velocities}

In Fig.~8 we plot \Vexp\ vs. the rate of mechanical energy of the stellar wind,
$L_w$, calculated as 1/2 $\dot M$  v$_\infty ^2$ (see data in Table~2).
A very clear tendency is found showing that stars with 
large $L_w$ have nebulae with large \Vexp. 
The linear fit has a correlation coefficient R = 0.71,
 indicating that, despite the scatter, the trend is real and much of the acceleration of the nebulae around [WC] stars should be due to the wind mechanical energy. Unfortunately only one WLPN and 
three ordinary PNe of our sample have published $\dot M$ and v$_\infty$, 
nevertheless, as expected considering their low $L_w$, all of them lie in the  low \Vexp\ zone.

\begin{figure}[!t]
\includegraphics[width=\columnwidth]{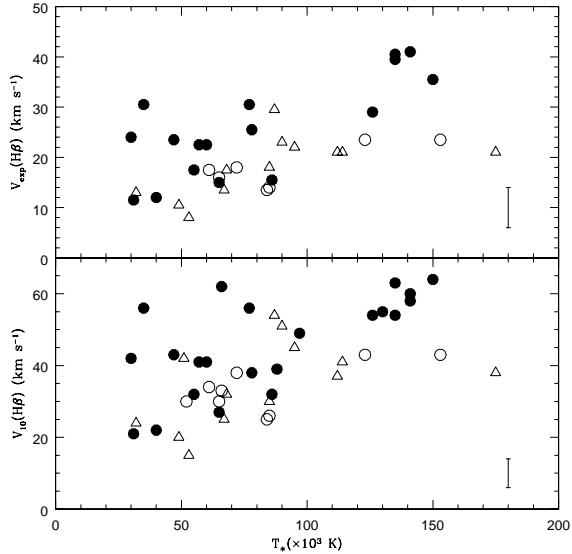}
\caption{\Vexp\ (top) and \Vdiez\ (bottom) are plotted versus the stellar temperatures. Symbols are as described in Fig. 4. Despite the large scatter, velocities and temperatures are correlated. The correlations coefficients are 0.64 for the top figure and 0.55 for the bottom figure.}
\label{Fig.6}
\end{figure}
\begin{figure}[!t]
\includegraphics[width=\columnwidth]{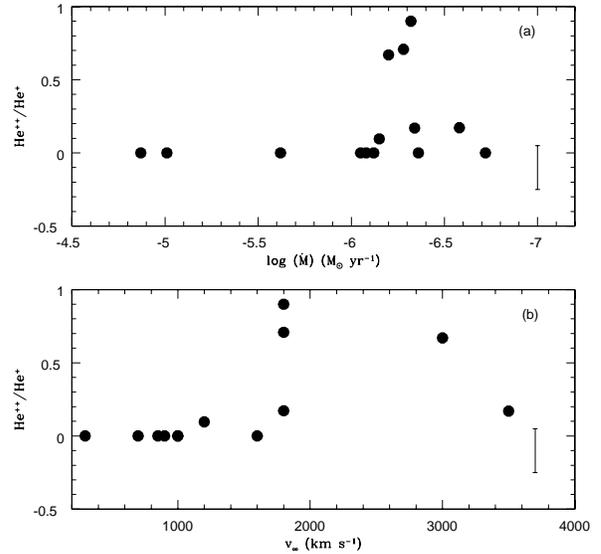}
\caption{Nebular ionization degree (He$^{++}$/He$^+$) is plotted against
some stellar parameters such as:
a) the mass-loss rate ($\dot M $), and b) the wind terminal velocity 
(v$_\infty$). Symbols are as described in Fig. 4.}
\label{Fig.7}
\end{figure}
\begin{figure}[!t]
\includegraphics[width=\columnwidth]{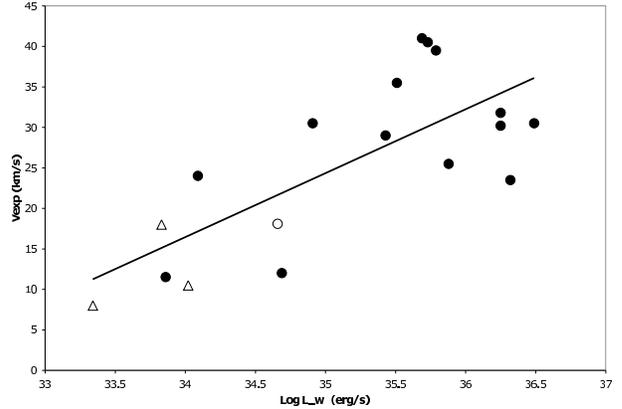}
\caption{\Vexp\ is plotted against the log of the rate of mechanical energy of the stellar wind, $L_w$. 
Symbols are the same as in Fig. 4. It is evident the relation between both parameters. The linear fit is
 \Vexp=7.8(log $L_w$) $-$249.3 \kms, with a correlation coefficient R = 0.71, 
therefore the large \Vexp \ of WRPNe can be attributed to acceleration by the [WC] stellar wind.}
\label{Fig.8}
\end{figure}

\section{Main conclusions}
In this work we have analyzed high spectral resolution data to 
study the kinematical behavior of a large sample of
planetary nebulae around [WC] stars, as well as PNe
around WELS and ordinary PNe.  
The  main conclusions are as follows:
\begin{itemize}
\item{ From a spatio-kinematic model representing an ellipsoidal nebula we have established 
that \Vexp \ can be measured as half the peak-to-peak separation of nebular lines for well 
resolved nebulae observed through a centered slit. The HWHM for objects with single 
symmetrical profiles is also a good indicator of \Vexp, in particular 
for the cases of low turbulence.}
\item[$\bullet$] { Considering the above, we measured \Vexp\ of several ions, 
for the objects in our sample showing double-peak or single symmetrical profiles.
 For a number of objects (K\,2-16, PRTM\,1, M\,1-32, M\,2-43, He\,2-429,
He\,2-459, K\,3-61, NGC\,6578, and YC\,2-32) these values are presented for the first time.}
\item[$\bullet$] {For all kind of objects  (WRPNe and non-WRPNe) it is found that the 
velocities of the different ions are very similar but in general: 
$$V_{\mathrm{exp}}([\ion{N}{II}]) \geq  V_{\mathrm{exp}}([\ion{O}{III}]), \ V_{\mathrm{exp}}(H\beta).$$
This indicates that all kind of nebulae presents  acceleration of the external shells.}
\item[$\bullet$] {We have shown that WRPNe have a different kinematical behavior than 
non-WRPNe.  For WRPNe \Vexp\ extends over a 
wide range from 10~\kms \ up to almost 50~\kms, while for ordinary PNe and WLPNe \Vexp \ is
always lower than 30~\kms. Average expansion velocities, measured from the peak-to-peak separation 
of split lines or from the FWHM of lines, are  40--45\% larger in WRPNe than in non-WRPNe. This
difference is larger for objects showing double-peak profiles. 
In addition to larger \Vexp, WRPNe seem to present more  turbulence than ordinary PNe.
These results are based mainly on our sample with double-peak profiles for which \Vexp \ and
turbulence effects are easily separated.}
\item[$\bullet$] {It is worth to remark that PNe around WELS were found to have a velocity
field very much alike the one of ordinary PNe, rather than the one of  WRPNe.  This
evidence shows that WELS,  contrary to the suggestion by Parthasarathy et~al.\@ (1998)}, 
do not belong to the [WC] family.
\item[$\bullet$] {Searching for evidence of high velocity material and perturbations of the
velocity field, half the widths at 1/10 of maximum intensity of the nebular lines 
(\Vdiez) were measured.  We have found that $\langle$\Vdiez$\rangle$ for WRPNe are also about 
40--45\% larger  than the averages for non-WRPNe. But in particular \Vdiez\ is large for some 
young compact WRPNe and  a few ordinary PNe which would have large turbulence or high 
velocity material in the central zones. This phenomenon is present in many planetary nebulae irrespective of their evolutionary age.}
%
%
\item[$\bullet$] {Strong evidence indicating that the nebular expansion 
increases with the nebular age (indicated by the nebular electron density),
regardless of the nebular type (WRPNe or non-WRPNe) has been found.}
\item[$\bullet$] { The same is found when analyzing the temperature of stars: the older the 
central star  (as measured by its  temperature), the faster the nebular expansion.  This
behavior occurs in non-WRPNe as well, but these objects seem to accelerate
to a maximum \Vexp\ of about 25--30 \kms\ when their stars reach $T_\ast$ of about 90,000--100,000~K, while WRPNe continue  accelerating up to reach \Vexp $\sim$~40 \kms.  The acceleration of nebulae is
 closely related to the mechanical energy of the stellar winds.
 [WC] winds accelerate more and apparently for longer time their nebulae. This constitutes an indirect evidence supporting the validity of the proposed 
evolutionary path:  [WC]-late stars $\rightarrow$ [WC]-early stars.  Another evidence for this
was established  years ago through the determination of the chemical abundances in [WC] stellar atmospheres. 
The results showed that C abundances were similar in [WC]-early and [WC]-late stars (De Marco 2002). 
However these results have been recently questioned by Hamann et al. (2005) on the base of improved non-LTE atmosphere models. These models show lower C abundance in apparently evolved [WC]-early stars than in less evolved [WC]-late ones, while one should expect the opposite. A further revision of this subject is certainly worthy.}
\end{itemize}

\acknowledgements
Invaluable comments by M.  Peimbert are deeply appreciated.  The authors are grateful 
to the staff members at OAN-SPM for their technical support. This work was
partially supported by DGAPA/UNAM (grants IN114601 and IN118405),
CONACYT/Mexico (grant 32594-E), and CONACYT-CNRS exchange agreement.  S.M.
acknowledges scholarship by DGEP/UNAM and CONACYT. M.P. is grateful to DAS, 
University of Chile for hospitality during a sabbatical visit, supported by FONDAP-Chile and DGAPA/UNAM. 


\bigskip

\begin{appendix}

\section*{APPENDIX A}

\subsection*{A. Deblending Nebular from Stellar Lines}
In Fig. 9 we present observed profiles of 
H$\alpha$, and [\ion{N}{II}] $\lambda\lambda$6548,6583 lines, and [\ion{S}{II}]  
$\lambda\lambda$6717,6731 for  different [WC]-type objects. Some stellar lines (usually very wide) appear at the same wavelength that the nebular lines and could affect 
the nebular line measurements if the spectral resolution is not adequate. These figures show 
that our  resolution of about  18,000, allows to safely deblend the stellar and nebular 
lines, even in the cases of [WC]-late stars whose stellar winds show velocities of a few 
100 \kms \ and thus their lines have FWHM of a few \AA. In a few extreme cases, where [\ion{N}{2}] $\lambda$6583 appears severely blended with the stellar \ion{C}{2} lines, as the case of M\,4-18 or NGC\,40, the 
line width of [\ion{N}{2}] $\lambda$6548 was measured.

\end{appendix}

\begin{appendix}

\section*{APPENDIX B}

\begin{figure*}[!th]\centering
     \begin{tabular}{llll}
       \includegraphics[width=4.5cm,height=4.5cm]{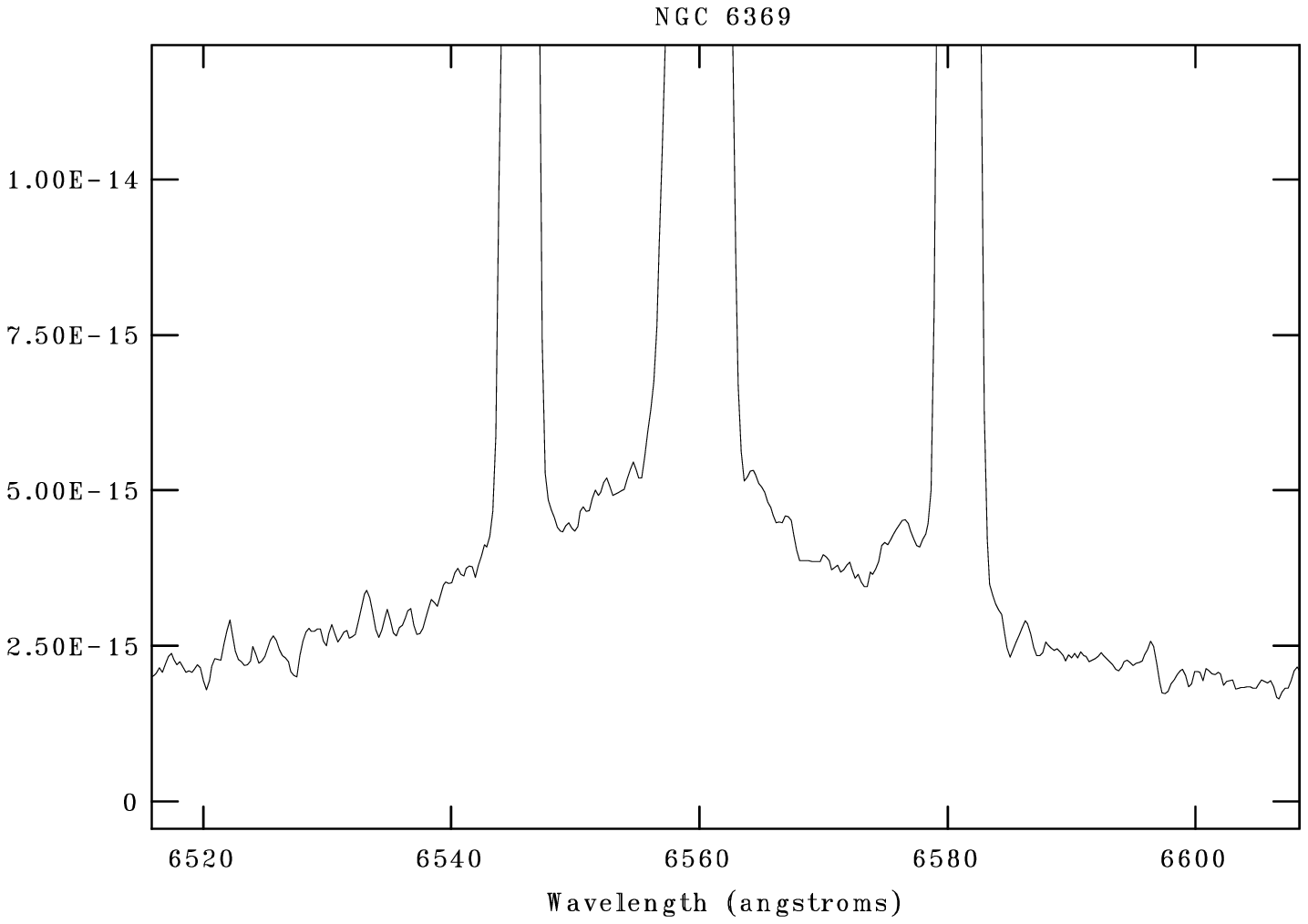} &
       \includegraphics[width=4.5cm,height=4.5cm]{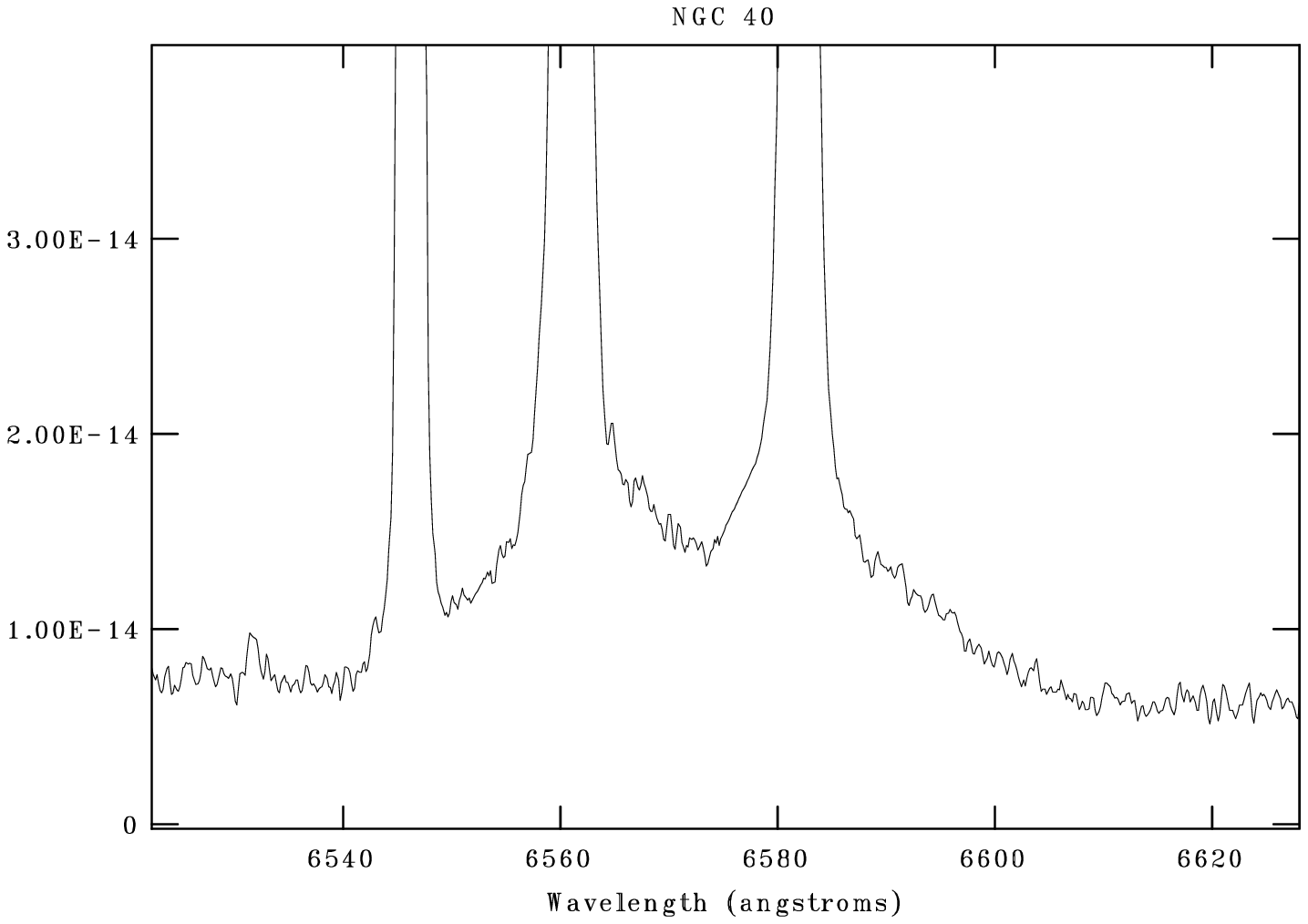} &
       \includegraphics[width=4.5cm,height=4.5cm]{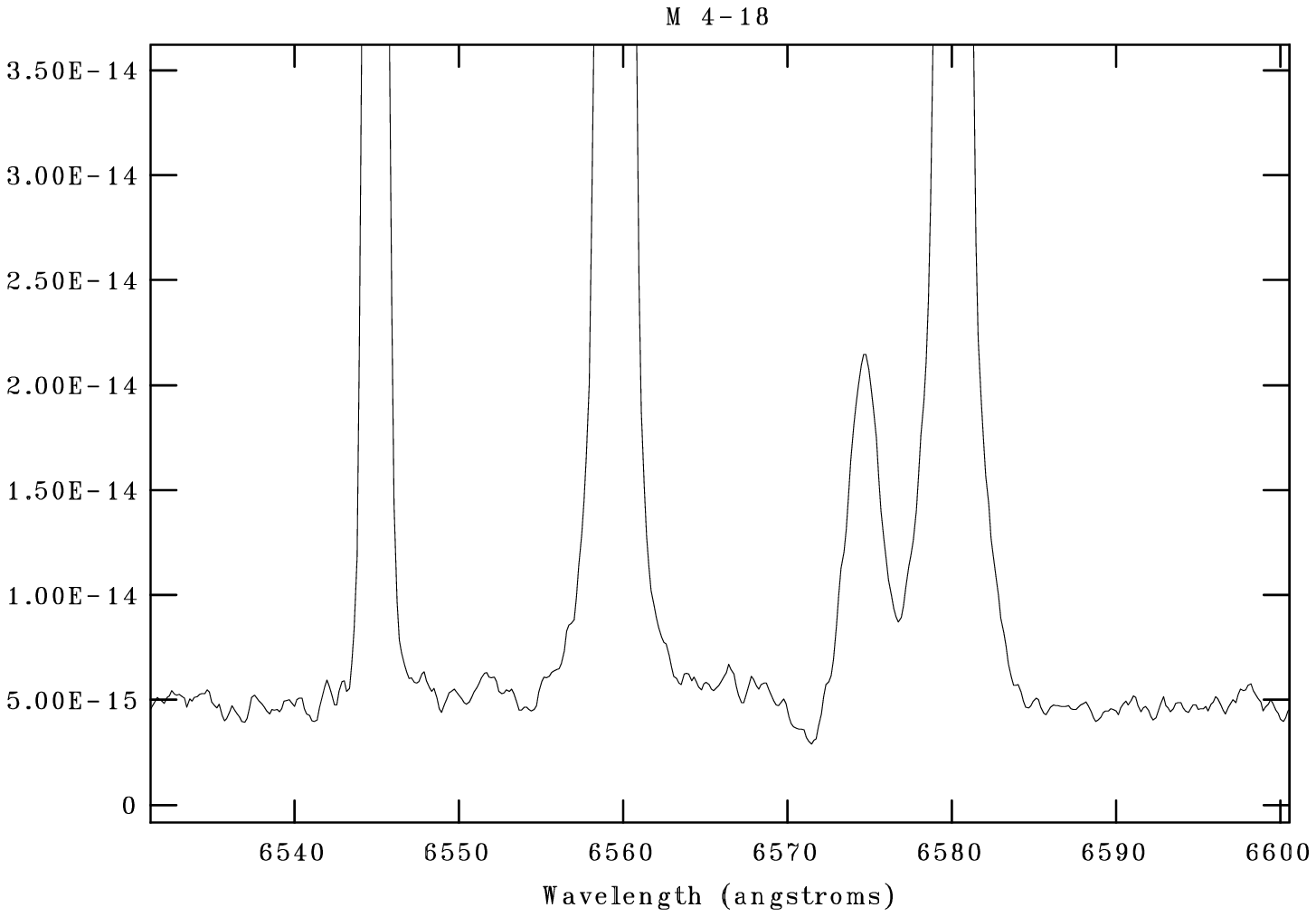}\\
       & \\
\includegraphics[width=4.5cm,height=4.5cm]{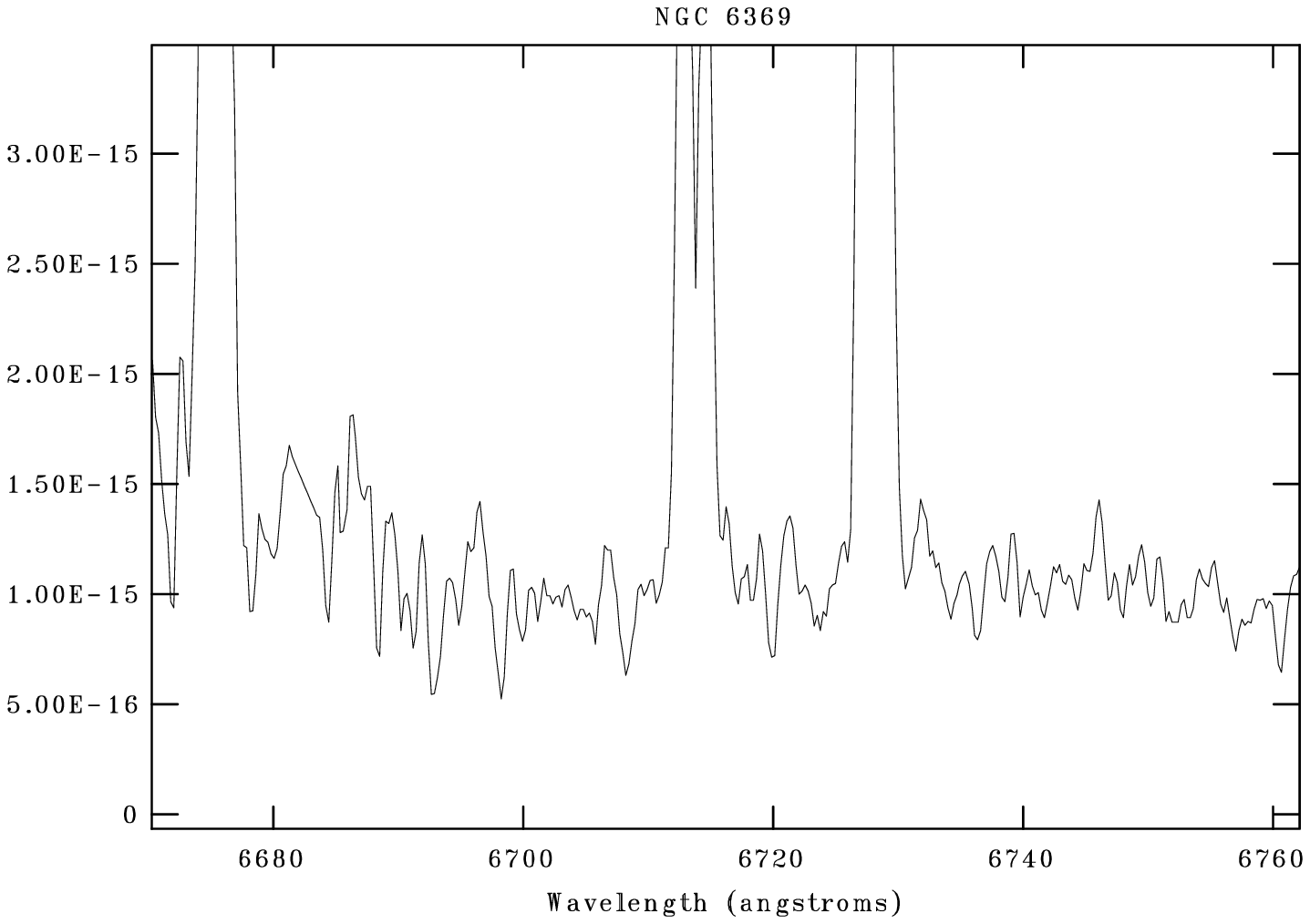} &
\includegraphics[width=4.5cm,height=4.5cm]{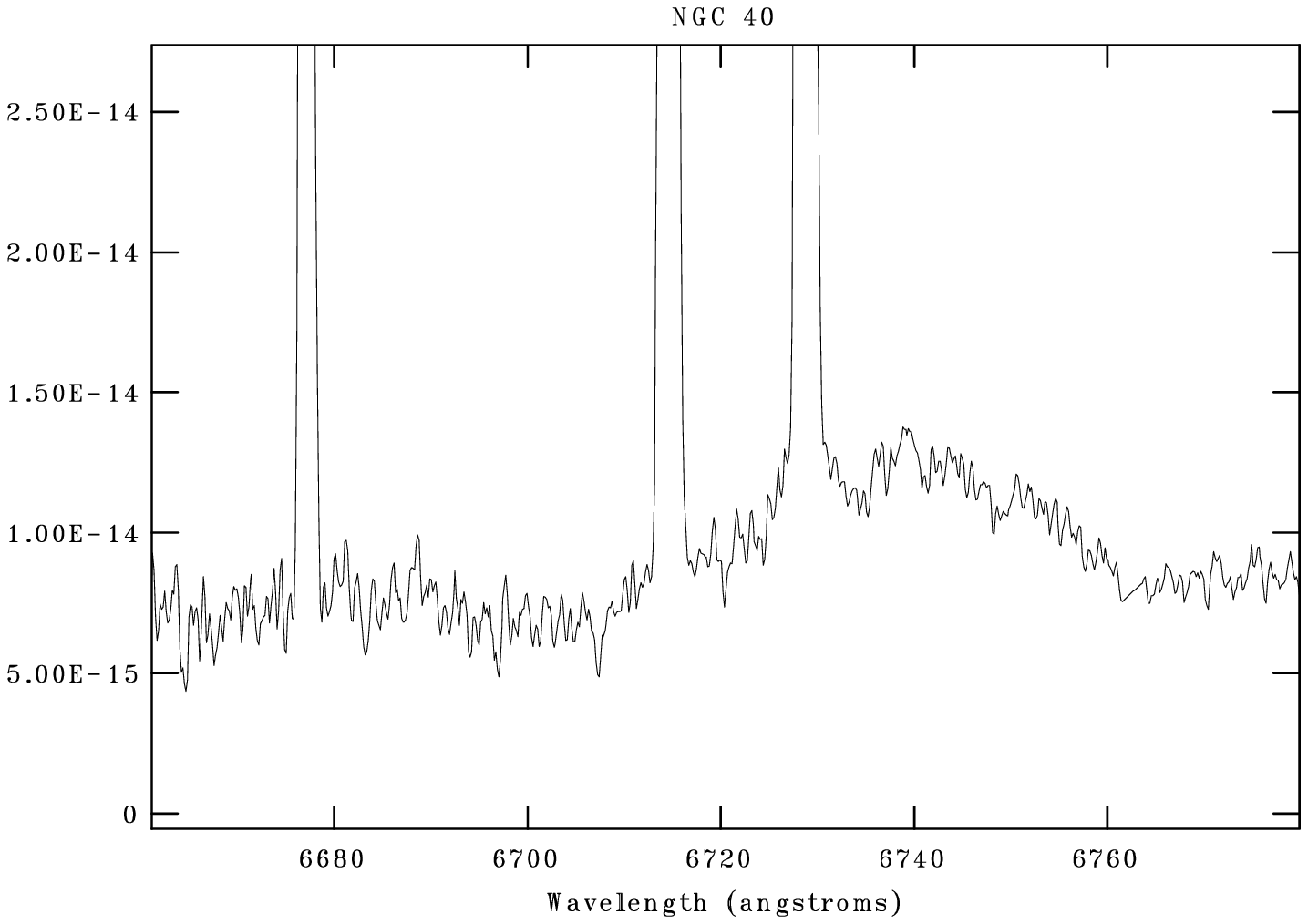} &
    \includegraphics[width=4.5cm,height=4.5cm]{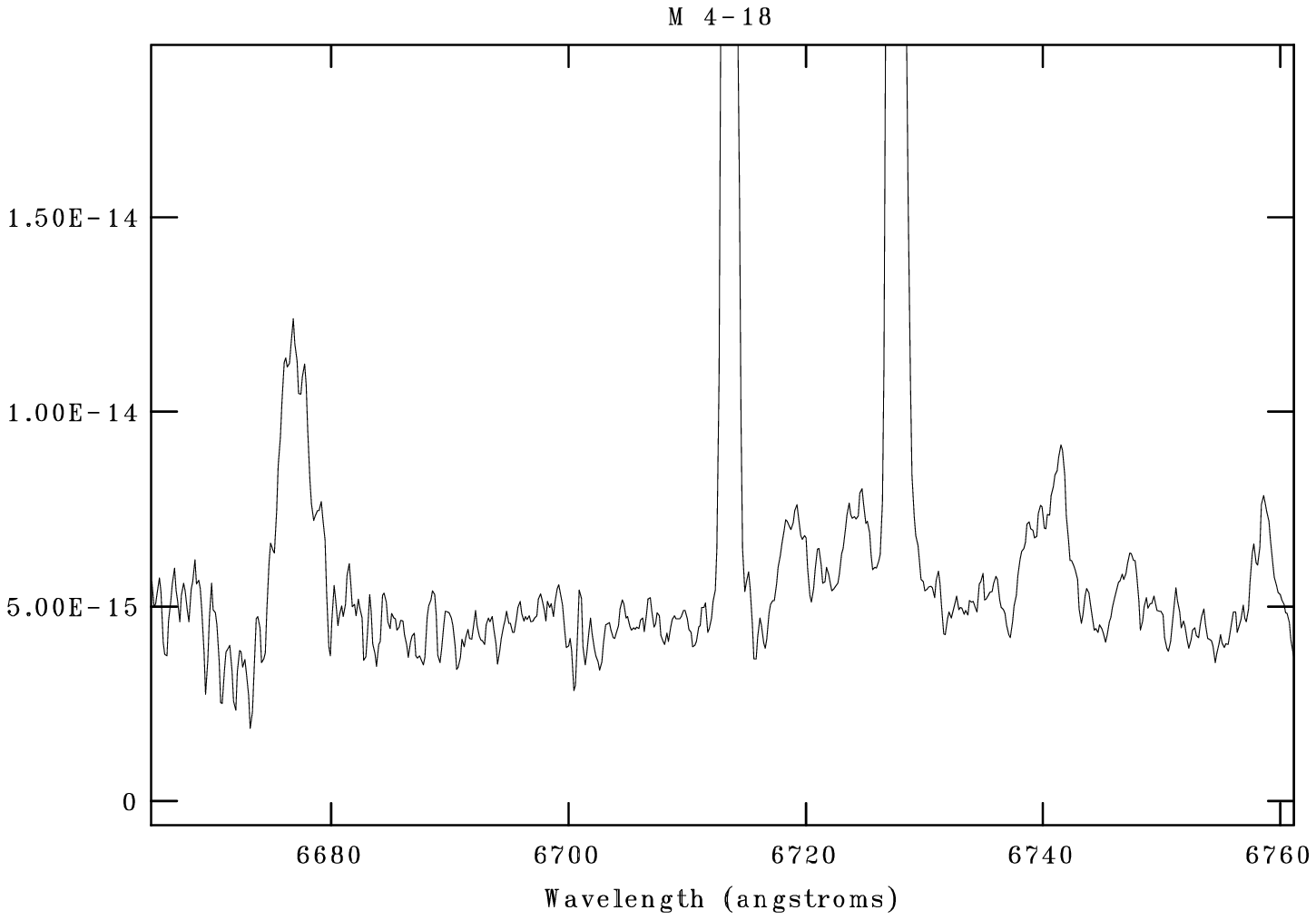} 
 \end{tabular}
   \caption[]{Profiles of H$\alpha$, and [\ion{N}{II}] $\lambda\lambda$6548,6583 lines, and [\ion{S}{II}] $\lambda\lambda$6717,6731, showing the nebular and stellar lines for different [WC]-type objects ([WC\,4] NGC\,6369, [WC\,8] NGC\,40, [WC\,10] M\,4-18. The nebular lines can be safely deblended from the wide stellar lines even in the cases of slow stellar winds as NGC\,40 and M\,4-18, except for [\ion{N}{2}] $\lambda$6583. [\ion{N}{2}] $\lambda$6548 can be measured instead.}
\end{figure*}
 
\centerline{B. PROFILE OF INDIVIDUAL OBJECTS}

\medskip

\centerline{B.1. {\it The Peculiar M\,1-32 and BD+30\degr3639}}

\smallskip

The WRPNe M\,1-32 and BD+30\degr3639 show single  symmetrical profiles, but 
they present \Vdiez\ from [\ion{O}{III}] larger than \Vdiez\ from \Hbeta\ by about 30\,\kms.  For  
BD+30\degr3639 this is also the case for  \Vexp. It is well known that this object shows this anomaly (Bryce \& Mellema 1999 and references therein) and we have 
found that M\,1-32 presents a similar behavior.  

The spatially resolved study of BD+30\degr3639 by Bryce \& Mellema
(1999) shows that the N$^+$ shell is more spatially extended but less spectrally
extended than the O$^{++}$ shell.  The [\ion{N}{II}] velocity ellipses presented by
these authors appear to be almost open-ended, with [\ion{O}{III}] emission emerging
from the gaps.  Bryce \& Mellema measured the expansion velocities from [\ion{N}{II}] 
and [\ion{O}{III}] profiles finding 28$\pm$1 and 36$\pm$1 \,\kms, respectively, 
which coincide within uncertainties with our values of 23 and 43\,\kms \ for the same ions. 
Acker et~al.\@ (2002) reported \Vexp \ of 27\,\kms \ with a
turbulent velocity of 15\,\kms \ for this object.

M\,1-32 has very unusual profiles characterized by a very-narrow intense component (with
FWHM smaller than 25\,\kms) and very wide wings, with a full width at the base of
about 125\,\kms \ (Fig.~3c).  The wings are much wider in [\ion{O}{III}]\,$\lambda$5007 
 and \ion{He}{I}\,$\lambda$5876 than in H$\beta$ and they
should be produced by high velocity gas in the unresolved central zone. Turbulence also 
can produce large wings, but a huge turbulence velocity should be required to create 
the wide wings observed (Morisset, in preparation). 

Both objects are low ionization nebulae (BD+30\degr3639
in particular), where O$^{++}$ lies in the inner nebular zone which is
probably interacting with (and being disturbed by) the [WC] 
wind.  It seems evident that the mechanical energy of the wind is strongly
affecting the inner zone in both nebulae.

\medskip

\centerline{B.2. {\it Objects Showing Single  Asymmetrical Profiles}}

\smallskip

 Fig. 3b shows the single asymmetrical profile of [\ion{N}{II}]\,$\lambda$6583 
for Hb\,4.  This is a high excitation Type~I planetary nebula
with a bright nebular core and outer low ionization ansae
(Corradi et~al.\@ 1996; Gon{\c{c}}alves et~al.\@  2001). L\'opez et al. (1997)
showed that the ansae are produced by collimated outflows with radial velocities of
$\pm$150~\kms, relative to the central core.  They measured an expansion velocity
of 21.5~\kms \ for the core. The model by Acker et~al.\@  (2002)
indicates expansion and turbulence velocities of 16 and 14~\kms, respectively.
Our \Vexp\ value for H$\beta$ is 16~\kms.

Our position-velocity diagram for the central core (Fig.~3b) shows a structure
similar to a broken or incomplete shell, with a bright compact knot in the red side
and fainter extensions to the blue.  This structure is more evident at
low-ionization species and produces asymmetrical single line profiles with a \Vdiez
\ of about 32~\kms.

Another nebula with similar asymmetrical profiles is IC\,1747, where we also
detected a bright knot and faint extensions to the blue.  Extracted profiles show
an intense component with an extended blue wing.  Images by Balick (1987) show a
knotty ring nebula.

The ordinary PNe M\,1-2, M\,2-53, and IC\,2149 show similar single asymmetrical
profiles.

\medskip

\centerline{B.3. {\it Objects Showing High Velocity Extensions}}
 \centerline{\it or Very High Turbulence}

\smallskip

The most conspicuous object presenting high velocity extensions is M\,1-32, described above.  Other objects showing the same phenomenon are:
He\,2-459, although the profiles are not as wide as in M\,1-32, and M\,3-15, where two very
faint high-velocity knots (located in the central zone) at $-90$ and 80~\kms \ are
detected.  In the latter case, the knots are faint with intensities below ${{1} \over
{10}}I_{\mathrm{max}}$; thus, only the velocity of the bright component is reported in
Tables~4 and 5.  M\,3-15 is also peculiar in the sense that this nebula shows an
extraordinarily low Ne abundance (Paper~II).

There are three of ordinary PNe showing very large \Vdiez.  
They are PRTM\,1, NGC\,4361, and M\,1-1. The first two are considered halo PNe,
and are both density bounded nebulae. The low ionization species are not detected in these objects (Pe\~na et~al.\@  1990; 
Torres-Peimbert et al. 1990). V\'azquez et~al.\@ (1999) reported high velocity 
material and a complex kinematics in NGC\,4361.  Similarly M\,1-1 shows a very high excitation 
with faint emission of the low ionized species. It is usual to find fast \Vexp\ in density bounded 
objects (Acker et~al.\@ 2002). In these thin low-mass nebulae the ionization front has broken 
through the outer layers which apparently expand fast having no neutral material around. 

\medskip






\centerline{B.5. {\it K\,2-16: a [WC]-Late Object}}
 \centerline{\it with Double-Peak Profiles}

\smallskip

\begin{figure}[!t]
\includegraphics[width=\columnwidth,height=\columnwidth]{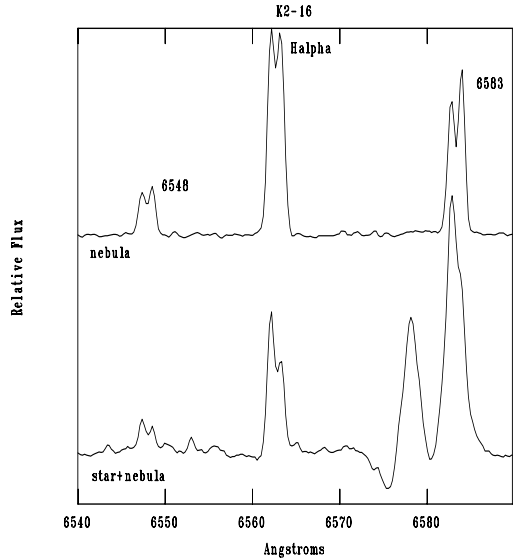} 
\caption{Nebular only (up) and nebular + stellar (low) extracted spectra in the zone of 
H$\alpha$ for K\,2-16. The nebular spectrum was extracted at about 4-5$''$ from the central
 star. The nebular lines show double-peak. The stellar + nebular spectrum shows the intense stellar 
 lines \ion{C}{2} lines well mixed with the nebular emission, preventing to measure [\ion{N}{II}] $\lambda$6583.}   
\label{fig.10}
\end{figure}

\begin{figure}[!t]
\begin{tabular}{l}
\includegraphics[width=\columnwidth,height=\columnwidth]{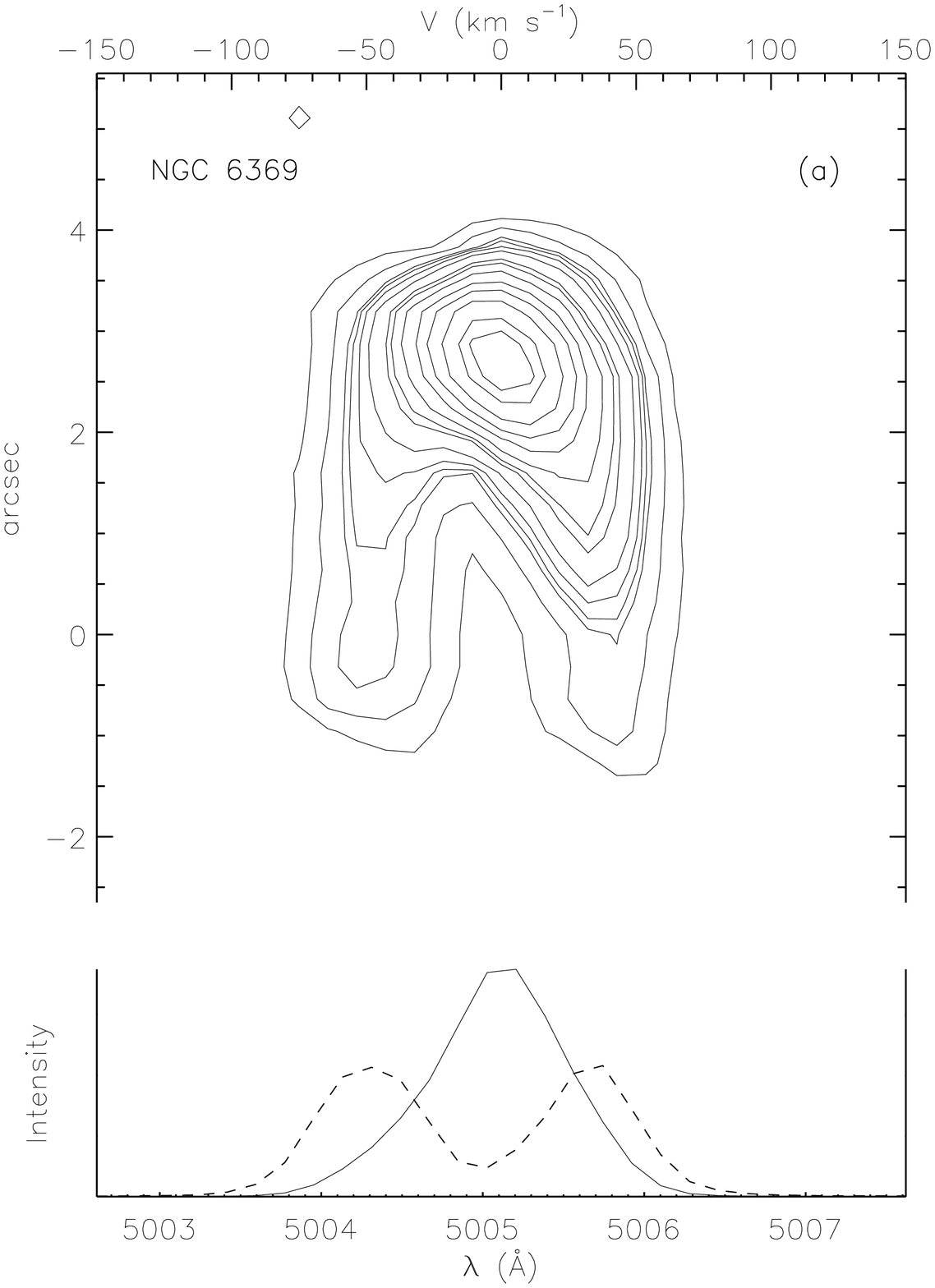} \\
\includegraphics[width=\columnwidth,height=\columnwidth]{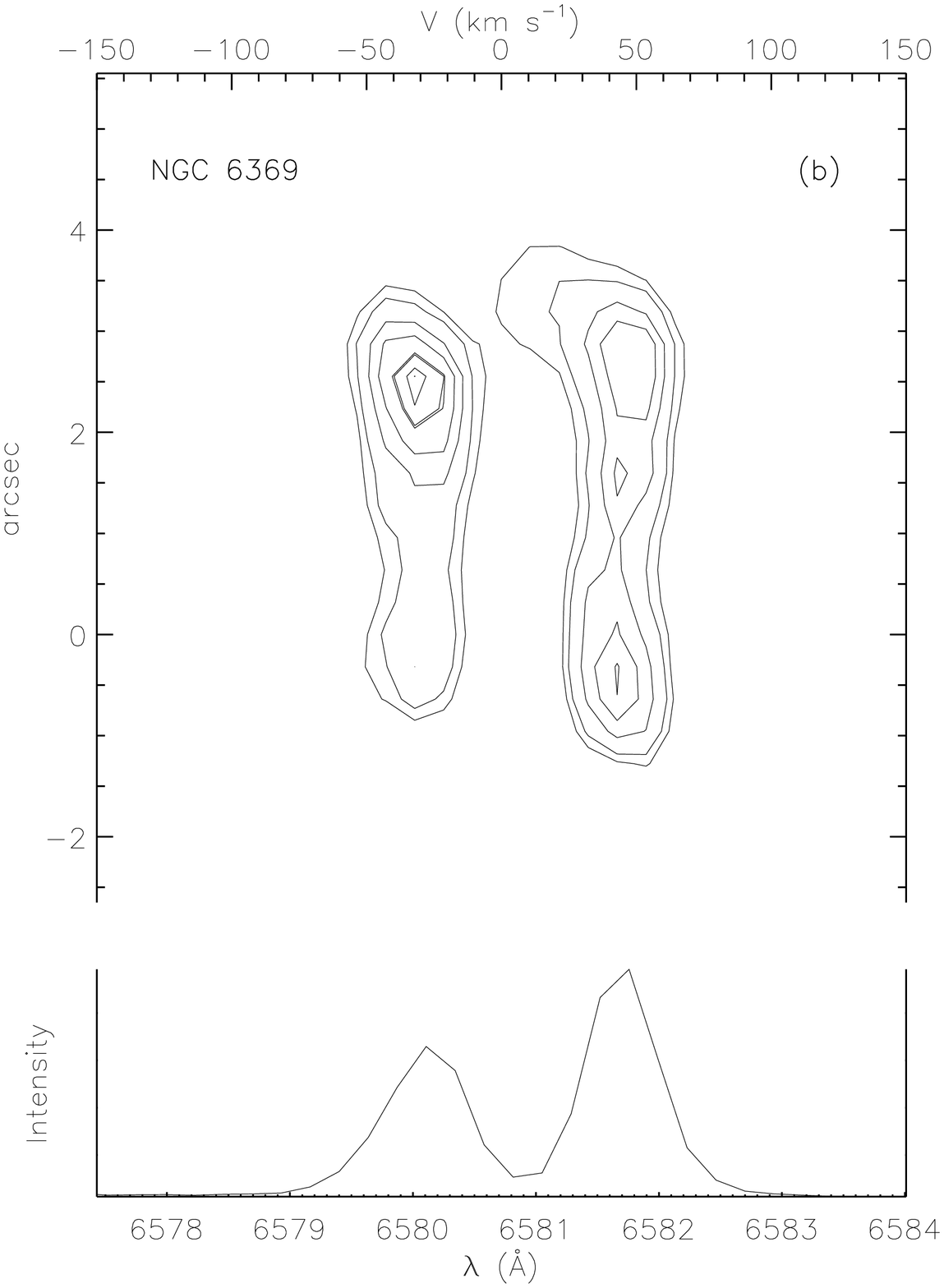} \\ 
\\
\end{tabular} 
\caption{Position-velocity contours and the extracted spectra from (a) 
[\ion{O}{III}]\,$\lambda$5007 and (b) [\ion{N}{II}]\,$\lambda$6583 are shown.  
The position of the central star is at 0\arcsec \ in both cases.  In (a), the solid line
corresponds to the emission of the bright knot at 3\arcsec \ East from the
central star, whereas the dashed double-peak line corresponds to the emission at the 
central star position (0\arcsec).  This emission is shown $5 \times$ its actual 
intensity to feature both lines in a single plot.  In (b), both emissions at 0\arcsec \
and 3\arcsec \ show the same split-line spectrum. The bright knot is not detected in [\ion{N}{II}]\,$\lambda$6583. }  
\end{figure}



This is a very-faint low-ionization PN with a shell morphology of 13$''$ in diameter.
  The shell structure as well as the low electron density  are unusual for a nebula
around a [WC\,11] star. In this sense K\,2-16 is similar to the also extended and low-density PM\,1-188, 
ionized by a [WC\,10] star (Pe\~na 2005).  K\,2-16 is the only nebula around a
[WC]-late star showing double peak profiles.  The star is very bright (V = 12.75~mag 
from Acker et~al.\@ 1992) and it presents intense emission lines severely contaminating 
some nebular lines, specially in the zone of [\ion{N}{II}] $\lambda$6583 and [\ion{S}{2}] $\lambda$6731.  Therefore we took special care in
measuring the nebular line widths in a zone at 4-5$''$ outside 
the stellar emission. In Fig.~10 we present the extracted spectra showing ``the nebular only'' and ``the nebular + stellar'' emissions.

   Our \Vexp\ value of 24~\kms \ is much smaller than the expansion
velocity determined by Acker et~al.\@  (2002), who proposed two models for this object: one
 with $V_{\mathrm{exp}} = 34$~\kms \ and a turbulent velocity of 12~\kms \ or alternatively, 
a model with $V_{\mathrm{exp}}= 38$~\kms \ and an acceleration producing a difference of 35~\kms\ 
between the velocities of the inner and outer radii.  
It seems that the stellar lines affecting the nebular [\ion{N}{II}] $\lambda$6583 has 
introduced an additional uncertainty in Acker et~al.  model.

The kinematics of K\,2-16 could be explained by an old thin shell
strongly accelerated by the [WC] wind of this apparently low-mass slowly-evolving star. 
 A very simple calculation to support this idea can be made by assuming that the nebular
 mass of K\,2-16 is about 0.1~$M_\odot$. Then the mechanical energy in the shell would be about 10$^{44.76}$~erg. 
Thus the wind (with mechanical luminosity  $L_w$=10$^{34.09}$~erg s$^{-1}$ as 
calculated from Table~2) should have been blowing for about 1500~yr in  order to provide 
the energy in the shell (an efficiency of a 100\% is assumed). A similar calculation for M\,4-18, 
gives an age of about 580~yr. In conclusion K\,2-16 should be several times older than M\,4-18, 
for possessing such a low density, large diameter and high \Vexp.

\medskip

\centerline{B.6. {\it Peculiar Line Profiles in NGC~6369}}

\smallskip

This extended nebula shows a peculiar position-velocity structure (see 
Fig.~11).  A beautiful well resolved picture of this object is found at
 http://heritage.stsci.edu/2002/25/index.html. The nebula consists of a faint central zone, 
a very bright extended ring, faint filamentary extensions  and ansa-type structures 
that were studied by  Gon{\c{c}}alves et al. (2001). Monteiro et~al.\@ (2004) modeled 
this object as a clumpy hourglass shaped nebula. In the $HST$ picture, the ionization 
structure is easily seen showing that the low ionization species are located mainly in the outer zone. 

In Fig. 11 we show that, at the central zone (0\arcsec, where the central star is located) 
well split lines for all the ions are detected, with \Vexp\ of about 36\,\kms.  
 At about 3--4\arcsec \ to the East from the central star we detect a very bright zone 
(corresponding to the inner side of the ring) 
  emitting mainly in high excitation lines (\ion{H}{I}, [\ion{O}{III}], \ion{He}{I}, and 
\ion{He}{II}). These lines present single profiles. The low ionization lines ([\ion{N}{II}], 
[\ion{O}{II}], [\ion{S}{II}] are weak here and they present double peak.   This 
should be the zone where Acker et~al.\@ (2002) obtained their spectrum with H$\alpha$ showing a 
single component and [\ion{N}{II}]\,$\lambda$6583, a complex profile (see their Fig.~2). Due to the complexity of the profile in this zone they could not model this object.

\end{appendix}

\end{document}